\documentclass[prd, twocolumn,
groupedaddress,nofootinbib,showpacs,amsfonts]{revtex4-1}
\usepackage{amsmath}
\usepackage{amssymb}
\usepackage{latexsym}
\usepackage{amsfonts}
\usepackage{epsfig}
\usepackage{psfrag}
\usepackage{graphicx}
\usepackage{subfigure} 
\usepackage{color}      
\usepackage[colorlinks=true, citecolor=blue, urlcolor=blue]{hyperref}

\newcommand{\lb}{{\langle}}
\newcommand{\rb}{{\rangle}}

\newcommand{\beq}{\begin{equation}}
\newcommand{\eeq}{\end{equation}}
\newcommand{\nn}{\nonumber\\}

\newcommand{\bea}{\begin{eqnarray}}
\newcommand{\ea}{\end{eqnarray}}

\begin{document}
\begin{abstract}
We studied the process of decoherence induced by the presence of an environment in acoustic black holes, using the open quantum system approach, thus extending previous work. We focused on the ion trap model but the formalism is general to any experimental implementation. We computed the decoherence time for that setup. We found that a quantum to classical transition occurs during the measurement and we proposed improved parameters to avoid such a feature. We provide analytic estimations for both zero and finite temperature. We also studied the entanglement between the Hawking-pair phonons for an acoustic black hole while in contact with a reservoir, through the quantum correlations, showing that they remain strongly correlated for small enough times and temperatures. We used the stochastic formalism and the method of characteristic to solve the field wave equation.
\end{abstract}

\title{Dynamics of an Acoustic Black Hole as an Open Quantum System}

\author{Fernando C. Lombardo and Gustavo J. Turiaci} 
\affiliation{Departamento de F\'isica Juan Jos\'e Giambiagi, FCEyN UBA, Facultad de Ciencias Exactas y Naturales, Ciudad Universitaria, Pabell\'on I, 1428 Buenos Aires, Argentina - IFIBA }                        

\date[]{August 1, 2012}
\maketitle

\section{Introduction}\label{intro}

Hawking effect, i.e. the particle creation process that gives rise to a thermal spectrum of radiation outgoing from a black hole \cite{hawking}, is a 
prediction of quantum field theory in curved space-time. This effect together with its entropy complete the interpretation of black holes as thermal objects. On one hand, it is believed that the heart of a theory that unifies quantum mechanics and gravity lies in understanding the nature of this thermality. On the other hand, it is also important to collect experimental evidence in order to gain insight into this phenomenon, but this is practically impossible since black holes' temperatures are less than $nK$. Since the phenomenon has not been observed experimentally, it is of crucial importance that all the assumptions underlying the Hawking effect be carefully studied so as to try to understand the process. 

W. G. Unruh proposed an analogue gravity hydrodynamical model where phonons propagate in a fluid with a subsonic and supersonic regime \cite{unruh-prl}. This model obeys the dynamics of a massless scalar field near a black hole and provides a possible experimental implementation  to study the Hawking effect. Subsequently, there have been several more realistic proposals that involved BEC \cite{bec}, moving dielectrics \cite{dielectric}, waveguides \cite{nonlinear}, slow light systems \cite{not-slow}, among others. 

On general grounds, an acoustic black hole is a system in
which phonons are unable to escape from a fluid
that is flowing more quickly than the local speed of sound. Therefore, trapped phonons
are analogous to light in real gravitational black holes. The phonons propagating in perfect fluids
exhibit the same properties of motion as fields, such as gravity, in
space and time.

 The current proposals do not provide conclusive evidence of the Hawking effect (see for example \cite{quilombo}). Nevertheless, we believe that a particular one by Horstmann \emph{et al.}, \cite{cirac}, provides a promising setup. This system consists of a circular ring of trapped ions moving with an inhomogeneous velocity profile emulating a black hole, as explained above. The signature of this quantum radiation is the correlation between entangled phonons near the horizon, \cite{corrbec}, which can be measured by coupling the ions' motional degrees of freedom to their internal state, \cite{medicion}. 

Given that we are interested in the quantum nature of the effect, in this paper we make a study of the interactions of the acoustic black hole with an environment and how this induces decoherence in the system.

The presence of an environment can destroy all the traces of the quantumness of a system. All real world quantum systems interact with their surrounding environment to a greater or lesser extent. As the quantum system is in interaction with an environment defined as any degrees of freedom coupled to the system which can entangle its states, a degradation of pure states into mixtures takes place. No matter how weak the coupling that prevents the system from being isolated, the evolution of an open quantum system is eventually plagued by nonunitary features like decoherence and dissipation. Decoherence, in particular, is a quantum effect whereby the system loses its ability to exhibit coherent behavior. Nowadays, decoherence stands as a serious obstacle in quantum information processing. All in all, it should be crucial to control decoherence in order to plan a concrete setup for measuring the Hawking radiation 
as a quantum feature of analogue gravity models. 

In this paper, we show a more detailed and deeper treatment of the problem given in \cite{tesis}, and we present new results concerning the evaluation of the decoherence time and analytical expressions for correlation functions of the quantum fields. We continue using the field-theoretical description in order to present a derivation applicable to any implementation of an acoustic black hole. Nevertheless, given our interest in the ion trap we provide numerical results corresponding to that particular setup.

The article is organized as follows: in Sec. \ref{sec:modelo} we show the specific model we are using. In Sec. \ref{sec:entorno} we present the model of 
environment and show the non-equilibrium dynamics of the system. Sec. \ref{sec:tdecc} contains the estimation of the decoherence time and we show, 
in Sec. \ref{sec:ep} the dynamics of the entanglement through the analysis of correlations. Finally, we summarize our results in Sec. \ref{sec:conc}.

\section{A particular Acoustic Black Hole: The Ion Ring}\label{sec:modelo}
As stated above, we will work in the context of the field theoretical description. Given that we are interested in the ion ring model, and with the purpose of introducing the subject of analogue gravity with a specific example, we will explain in detail this set up from first principles. We will begin with the discrete description of $N$ ions of mass $m$ in a circular trap of radius $R$ to end with the description of a massless scalar field in a curved background.
%\begin{figure}[h]
%\begin{center}
%\subfigure[Anillo de Iones]{
%\includegraphics[width=70mm]{figure1.pdf}}
%\subfigure[Perfil de Velocidad]{
%\includegraphics[width=70mm]{figure2.pdf}}
%\end{center}
%\caption{\label{arreglo} Schematic setup for the ions inside the circular trap. The required forces to impose the velocity profile are localized in the transition intervals. The black region corresponds to the acoustic black hole inside the supersonic region. [Extracted from \cite{cirac}].}
%\end{figure}

Following the study done in Ref. \cite{cirac}, the Hamiltonian describing the ions of the circular trap is given by 
\begin{equation} \label{calz}
\mathcal{H}=-\sum_{i=1}^{N} \frac{\hbar^2}{2mR^2}\frac{\partial^2}{\partial\theta_i^2} + \sum_{i=1}^{N}V^e(\theta_i,t) +V^c(\theta_1,...,\theta_N),
\end{equation}
where $V^e(\theta_i,t)$ is an external field potential corresponding to an electric field that induces the classical trajectories $\theta_i^0(t)$ and $V^c(\theta_1,...,\theta_N)$ is the Coulomb potential between the ions. Those are such that the velocity profile is 
\begin{displaymath}
   v(\theta,t) = \left\{
     \begin{array}{lr l}
       v_{\text{min}} &  &   0 \leq \theta \leq \theta_H - \gamma_1\\
        \beta+\alpha   \big(\frac{\theta-\theta_H}{\gamma_1}\big) &&  -\gamma_1 \leq (\theta-\theta_H) \leq  \gamma_1\\
       v_{\text{max}} &&   \theta_H+\gamma_1 \leq \theta \leq 2\pi-\theta_H - \gamma_2\\
        \beta-\alpha  \big(\frac{\theta-2\pi+\theta_H}{\gamma_2}\big) &&  -\gamma_2 \leq (\theta-2\pi+\theta_H) \leq \gamma_2\\
       v_{\text{min}} &&   2\pi-\theta_H+\gamma_2 \leq \theta \leq 2\pi,
      \end{array}
   \right.
\end{displaymath} 
where $\beta=(v_{\text{max}}+v_{\text{min}})/2$ and $\alpha=(v_{\text{max}}-v_{\text{min}})/2$. The minimum and maximum velocity are constrained by the condition that each ion has to make one revolution during a period $T$. It is important to notice that we use an approximate velocity profile, as introduced in Ref. \cite{tesis}. The real, as explained in Ref. \cite{cirac}, must be $C^3$ but ours is a good approximation appropriate to our calculations. Taking this into account, the parameters $\gamma_i$ and $\theta_H$ do not exactly match those of Horstmann \emph{et al}. The system may be prepared in an initial thermal state with temperature $T_0$. An illustration of the setup and the velocity profile can be found in Figs. 1 and 2 of Ref. \cite{cirac}.

Initially, the velocity profile is constant with $v_{\text{min}}(t=0)=v_{\text{max}}(t=0)=2\pi/T$ but during a time $\tau$ the profile change such that $v_{\text{min}}(t\gg\tau)<2\pi/T$ and $v_{\text{max}}(t\gg\tau)>2\pi/T$ in the following way
\begin{equation}
v_{\text{min}}(t)=v_{\text{min}}+\bigg( \frac{2\pi}{T}-v_{\text{min}}\bigg)e^{ -t^2/\tau^2 },
\end{equation}
where we call $\tau=0.05\cdot T$ the collapse time, which is much smaller than $T$, and after that time the asymptotic profile is given by the parameter $v_{\text{min}}(t\gg\tau)=v_{\text{min}}$. It is important to notice that the measurement can only be performed during one period $T$, since after that time the system's classical configuration loses its stability, see Ref. \cite{cirac}. Therefore, we assume that any possible measurement ought to end after one revolution, but must last longer than $\tau$ in order for the acoustic black hole to form, as we will see below.

The problem can be linearized for small perturbations of the trajectories $\theta_i=\theta_i^0+\delta\theta_i$ as
\begin{equation}
\mathcal{H}\approx-\sum_{i=1}^{N} \frac{\hbar^2}{2mR^2}\frac{\partial^2}{\partial\delta\theta_i^2} +\frac{m}{2}\sum_{i\neq j} F_{ij}(t)\delta\theta_i\delta\theta_j.
\end{equation}
We are interested in the continuous description of this system, described by the field $\Phi(\theta=\theta_i^0(t),t)=\delta\theta_i(t)$. After diagonalizing the previous hamiltonian and going to the continuous limit we end up with the Lagrangian
\bea
\mathcal{L}[\Phi]=\int_0^{2\pi}d\theta\frac{\rho(\theta)}{2}\bigg\{(\partial_t\Phi&&+v(\theta)\partial_{\theta}\Phi)^2\nonumber\\
&&-(D(-i\partial_{\theta})\Phi(\theta,t))^2\bigg\},
\ea
where $D(\theta,k)=c(\theta)k+\mathcal{O}(k^3)$ is the dispersion relation. The speed of sound is given by 
\begin{equation}
c(\theta)=\sqrt{\frac{2n(\theta)Q^2}{mR^3}},
\end{equation}
where $n$ is the local density of the ions and $Q$ their electric charge. The conformal factor is $\rho(\theta)=mR^2n(\theta)=mR^2(N/(v(\theta)T))$. 

In Ref. \cite{unruhrob}, Unruh has shown that the Hawking radiation is robust against small deviations from a linear dispersion relation, so we take simply $D(\theta,k)=c(\theta)k$. Moreover, the conformal factor can be included to first order in the definition of the field, redefined as $\phi=\sqrt{\rho}\Phi$. Taking this into account, we can describe the field by means of an action written in the suggestive form (this description is the starting point also in Ref. \cite{tesis}) 
\begin{equation}
S[\phi]=\frac{1}{2}\int d^2x\sqrt{-g}g_{\mu\nu}\partial^{\mu}\phi\partial^{\nu}\phi,  \label{intro:ac}
\end{equation}
where $x^{\mu}=(t,x=\theta)$ and the effective metric is 
\beq\label{g}
ds^2=(c^2-v^2)dt^2+2vdxdt-dx^2.
\eeq

Of course this system is non-relativistic and it must be thought of as an effective sigma-like model describing the phonons in the ion trap. Nevertheless, this is indeed an analogue black hole, since it presents an event horizon in the points where $v^2=c^2$. This analogy can be clarified if we change the variables
$$
t \rightarrow \tau = t + \int \frac{v}{c^2-v^2}dx,
$$
then the metric is 
\beq\label{sch}
ds^2=(c^2-v^2)d\tau^2-\frac{dx^2}{1-v^2/c^2},
\eeq
and expanding linearly around the angle $\theta_H$ of the event horizon where $v(\theta_H)=c(\theta_H)$, we recover the Schwarchild metric. This field $\phi(x)$ describing the phononic exitation of the continuous array of ions can now be quantized. Repeating the Hawking radiation derivation, the Hawking temperature can be obtained 
\beq
T_H=\frac{\hbar}{4\pi v k_B}\frac{d}{d\theta}(v^2-c^2)\bigg|_H. \label{hawtemp}
\eeq
%\begin{figure}[ht]
%\begin{center}
%\subfigure[~Momentum Correlation]{
%\includegraphics[scale=0.29]{figure8a.pdf}}
%\subfigure[~Evolution of the Logarithmic Negativity]{
%\includegraphics[scale=0.25]{figure13.pdf}}
%\end{center}
%\caption{\label{ciraccorr} (a) Plot of the momentum-momentum correlation. The peak corresponding to the entanglement between the Hawking pair is marked with a dashed line. (b) Evolution of the logarithmic negativity as a function of time for different temperatures. [Extracted from \cite{cirac}]}
%\end{figure}
As suggested by Balbinot \emph{et al.} \cite{corrbec} and studied by Sch\"utzhold and Unruh \cite{entre:unruh-cor}, the signature of the quantum radiation, in contradistinction with a stimulated emission, and what would be measured in the proposal, is the peak present in the correlation $\lb\delta p_L(\theta)\delta p_L(\theta')\rb$, between two points inside and outside the acoustic black hole. The magnitude $\delta p_L$ is the left moving component of the canonical momentum conjugate to $\delta \theta$.  

This feature reflects the entanglement between the Hawking pair of phonons emitted near the event horizon. Following the discussion in Ref. \cite{corrbec}, this can be seen writing the ``in" vacuum in terms of the ``out" vacuum as a squeezed state
\beq
|{\rm in}\rb \propto \exp{\bigg( \sum_{\omega} e^{-\hbar\omega/2k_BT_H}a_{\omega}^{(esc)\dagger}a_{\omega}^{(tr)\dagger}}\bigg)|{\rm out}\rb \label{squeezed}
\eeq
where $a^{(esc),(tr)\dagger}$ are creation operators for, respectively, the outgoing escaping and trapped modes.
 
This magnitude was calculated numerically and the results are shown in Fig. 9 of Ref. \cite{cirac}. When one increases the initial temperature $T_0$ of the system, the entanglement is lost due to a quantum to classical transition induced by thermal effects. In \cite{cirac} they computed, as a measure of the entanglement, the logarithmic negativity, see Fig. 14 of Ref. \cite{cirac}. On one hand, for fixed temperature, the system needs a certain amount of time to generate the entanglement through the creation of the Hawking pairs. As the temperature increase above a threshold, $\sim100\cdot T_H$, the entanglement is lost and the system starts behaving classically. This is also reflected in the correlation at the same temperature scale. In this paper we will be interested in the quantum to classical transition induced by the non-equilibrium dynamics with an environment instead.

Another feature that can affect the measurement is the presence of a noise in the force used to produce the trajectories. This stochastic component of the force can be described by the parameter $\gamma$ defined such that $({\rm noise})=\gamma\times({\rm mean~force})$. The requirement that the peak in the correlation is distinguishable imposes the following bound 
\beq
\gamma\lesssim 5\cdot 10^{-6}.
\eeq

Although the problem can be treated in a discrete fashion, we choose to stick to the field description since the action given in Eq. (\ref{intro:ac}) is common to every realization of acoustic black holes, not restricted to ion traps. For example, in the case of a BEC acoustic black hole it emerges from the Gross-Pitaevski equation or in the waveguide setup from Maxwell's equations, etc, but the point is that the system must always be described with an action as Eq. (\ref{intro:ac}) in some regime. Therefore, it is important to notice that the analysis carried in the following sections can be applied to any set-up.
\section{Characterization of Bosonic Environments and the Non-equilibrium Dynamics} \label{sec:entorno}
\subsection{Interaction Model}
Our aim in this paper is to study the behavior out of equilibrium of this system while interacting with a quantum environment, since this coupling acts as a mechanism that induces decoherence and the system starts behaving classically after the decoherence time. As usual for this kind of tasks we use the Schwinger-Keldysh or ``in-in" formalism.

As explained in the previous section, our system is described with the action of a massless scalar field in a dynamical background,
\beq
S_0[\phi]=\frac{1}{2}\int d^2x \sqrt{-g}g_{\mu\nu}\partial^{\mu}\phi\partial^{\nu}\phi.  \label{ac}
\eeq
Following the quantum Brownian motion (QBM) paradigm, the environment is described by a continuous array of bosonic quantum harmonic oscillators distributed in each position of the circular trap, following Ref. \cite{tesis}. The bath is at rest with respect to the laboratory and it will be represented by the following degrees of freedom $q_{\nu}(\theta,t)$ with the action
\beq
S_{\mathcal{E}}[q_{\nu}]=\frac{1}{2}\int_0^{\infty} d\nu I(\nu) \int d^2x   \big[ \dot{q}^2_{\nu}(x,t)-\nu^2 q_{\nu}^2(x,t)\big].
\eeq
The function $I(\nu)$ corresponds to the mass of each oscillator that compose the environment. The interaction between the system and the environment is given by the following term in the action
\beq
S_{\rm int}[\phi,q_{\nu}]=-\int_0^{\infty} d\nu\int \zeta(\nu) d^2x~\phi(x) q_{\nu}(x),
\eeq
in such a way that the total action is
\beq
S[\phi,q_{\nu}]=S_0[\phi]+S_{\mathcal{E}}[q_{\nu}]+S_{\rm int}[\phi,q_{\nu}].
\eeq

The nature of the environment depends strongly on the specific model of black hole considered. For example, in the ion ring proposal the velocity profile is generated by the action of an electric field, which in turn is produced by plane-parallel electrodes. The irregularities of its surface produce fluctuations in the force they induce. The noiseless component of this force is included in the effective action of the system in Eq. (\ref{ac}), such that the analogue gravity model is achieved. The pure noise component due to the fluctuations can be introduced as an environment and the oscillators represent the corresponding modes of the stochastic electric field. The details of the possible natures of an environment was extensively discussed in Ref. \cite{tdec:wineland}. The fact that we are thinking of an electric field coupled to the ions' coordinates justifies the bilinear coupling used here. Nevertheless, in other analogue gravity models as long as the environment is bosonic, this model is fairly general. For example, if the field is coupled to the ions' velocity instead of their position, then this amounts only to a change $I(\nu)\mapsto I(\nu)\nu^2$, as can be seen below. Therefore, the information of each particular case is encoded in the spectral density.

To simplify the analysis we take an ohmic environment, although more general environments do not change substantially the results as can be seen in Ref. \cite{tdec:entorno} in the context of quantum Brownian motion.

In Ref. \cite{miles} it was shown that the presence of an environment with scales bigger than the Planck scale can generate instabilities such as Miles-type instabilities that jeopardize the detection of the Hawking radiation as a quantum induced effect. The environment is at rest with respect to de laboratory and within the approximation used, both the environment as the ions are in the linear dispersion relation regime, far from being in the Planckian regime, where the correct description is the Coulomb chain expression. Therefore, the model used here does not present any instability whatsoever induced by the environment.

In the following we will present the formalism used to study the non-equilibrium dynamics of this composed system.
\subsection{Reduced Density Matrix}
The dynamics of any system out of equilibrium is described by its density matrix together with its evolution in time. The matrix elements of this operator are given by
\beq
\rho(\phi,q;\phi',q'|t)=\lb\phi,q|\widehat\rho(t)|\phi',q'\rb.
\eeq
We use an uncorrelated initial state between the system and environment, in such a way that
\beq
\hat{\rho}(t_i)=\hat{\rho}_{\mathcal{S}}(t_i)\otimes \hat{\rho}_{\mathcal{E}}(t_i).
\eeq

These initial density matrices correspond to a thermal state with temperature $T_0=(k_B\beta)^{-1}$, i.e. $\rho \sim e^{-\beta H}$. In principle the initial temperature of the system and the environment can be different and the result we obtain below for the decoherence depends only on the initial environmental temperature. Nevertheless, it is more appropriate in the experiment to think of an initial equilibrium state between the system and the bath.

More general states do not change substantially the process of decoherence, as can be seen in Ref. \cite{estadoinicial}.

The reduced density matrix, which represents the concept of coarse graining the environmental degrees of freedom, which are of no interest to us, is defined in the usual way as the partial trace over their degrees of freedom
\beq
\rho_{\rm r}(\phi,\phi'|t)={\rm Tr}_{\mathcal{E}}\rho.
\eeq
\begin{widetext}
The generalization of the non-equilibrium dynamics of a quantum harmonic oscillator to the study on fields is straightforward and can be found in Ref. \cite{lombardo}. The evolution in time of the reduced density matrix is given by the following expression
\beq
\rho_r(\phi,\phi';t)=\int \mathcal{D} \phi_i\mathcal{D}\phi'_i\mathcal{J}_r(\phi,\phi';t|\phi_i,\phi'_i;t_i)\rho_{\mathcal{S}}(\phi_i,\phi'_i;t_i), \label{t:rr}
\eeq
where we defined the evolution operator associated with the reduced density matrix
\beq
\mathcal{J}_r(\phi,\phi';t|\phi_i,\phi'_i;t_i)\equiv \int_{\phi_i}^{\phi} \mathcal{D}\phi\int_{\phi'_i}^{\phi'}\mathcal{D}\phi' e^{i(S_0[\phi]-S_0[\phi'])/\hbar}\mathcal{F}[\phi,\phi'].
\eeq
In this expression the influence functional is defined following the Feynman-Vernon treatment
\beq
\mathcal{F}[\phi,\phi']\equiv e^{iS_{\rm IF}[\phi,\phi',t]/\hbar}=\int_{\rm CTP}\mathcal{D}q_{\nu}~\rho_{\mathcal{E}}(q_i,q'_i)\exp{\bigg\{i(S_{\mathcal{E}}[q_{\nu}]-S_{\mathcal{E}}[q'_{\nu}]+S_{\rm int}[\phi,q]-S_{\rm int}[\phi',q])/\hbar\bigg\}},
\eeq
where the prime and unprimed represent both branches of the closed time path curve over which we integrate, see Ref. \cite{lombardo,calhu}.

Therefore, the evolution of the system is dictated by the coarse grained effective action
\beq 
S_{\rm eff}[\phi,\phi']=S_0[\phi]-S_0[\phi']+S_{\text{IF}}[\phi,\phi'].
\eeq
The computation of the Feynman-Vernon influence action is identical to the QBM case and the result can be cast into the form
\beq
S_{IF}[\phi,\phi']=\int d^2xd^2x'\phi^-(x)\bigg({\bf D}(x,x')\phi^+(x')+\frac{i}{2}{\bf N}(x,x')\phi^-(x')\bigg),  \label{if}
\eeq
with the usual definitions $\phi^-=\phi-\phi'$ y $\phi^+=(\phi+\phi')/2$. The kernels ${\bf D}(x,x')$ and ${\bf N}(x,x')$ have the same expression that the QBM case regarding the temporal behavior, and they are local in space.
\bea
{\bf D}(x,x')&=&\int_0^{\infty}d\nu J(\nu)\sin{\nu(t-t')}\Theta(t-t')\delta(x-x'),  \label{d}\\
{\bf N}(x,x')&=&\frac{1}{2}\int_0^{\infty}d\nu J(\nu)\coth{\frac{\beta\nu}{2}}\cos{\nu(t-t')}\delta(x-x'),\label{n}
\ea
where the effective spectral density $J(\nu)$ is defined as 
\beq
J(\nu)=\frac{\zeta^2(\nu)}{I(\nu)\nu}.
\eeq
\end{widetext}
As we stated previously, we consider an ohmic environment and this translate into the following form of the spectral density
\beq
J(\nu)=\tilde{\gamma}^2\nu f(\nu) \label{espectral}
\eeq
where $\tilde{\gamma}$ plays the role of an \emph{effective coupling constant} and $f(\nu)$ is a generic cut-off function whose effect is to regularize the expression, and to this aim it must satisfy the following requirements
\beq
f(\nu=0)=1~~~{\rm and}~~~ f(\nu\gg\Lambda)\rightarrow0.
\eeq

Starting with the effective action $S_{\rm eff}$, we can write the semiclassical equation of motion, analogous to the stochastic Langevin equation for the field, i.e.
\beq
\frac{1}{\sqrt{g}}\frac{\partial}{\partial x^{\mu}}\bigg(\sqrt{g}g^{\mu\nu}\frac{\partial\phi}{\partial x^{\nu}}\bigg)+\int ds{\bf D}(t,s)\phi(s,x)=\xi(x,t). \nn \label{eq:lang}
\eeq
The field $\xi$ is a stochastic force with a gaussian probability distribution with zero mean $\lb \xi \rb=0$ and two-point function given by
\beq
\lb\xi(x)\xi(x')\rb=\hbar {\bf N}(x,x'). 
\eeq
This stochastic force can be identified as the noise presented in the previous section, quantified with the parameter $\gamma$.
\subsection{Estimation of the Effective Coupling Constant}
To make contact with the parameters used in Ref. \cite{cirac} we will calculate the relationship between the effective coupling constant $\tilde{\gamma}$ coming from the environmental spectral density and the parameter that characterizes the noise, the fluctuations in the force applied to the ions, introduced above as $\text{noise}=\gamma \overline{F}$.

Making a comparison between the equation that defines $\gamma$ with the Eq. (\ref{eq:lang}) it is possible to identify the stochastic force with the fluctuations $\gamma \overline{F}$. Using the equation $\Box \phi + \ldots = \xi$ one can obtain $m\Box( R\delta\theta) + \ldots =mR\xi/\sqrt{\rho}$ and therefore
\beq
\frac{mR}{\sqrt{\rho}}\xi_{\rm rms}=\gamma\overline{F},
\eeq
where $\xi_{\rm rms}=\sqrt{ \lb \xi\xi \rb}$ is the root mean square of the stochastic force.

The mean force can be estimated as $\overline{F}\sim mR(v_{\text{max}}-v_{\text{min}})/\tau$. Therefore, coming back to the previous expression
\beq
\frac{1}{\sqrt{\rho}}\sqrt{ \lb \xi\xi \rb}= \gamma \frac{1}{\tau}(v_{\text{max}}-v_{\text{min}}).
\eeq

The noise kernel is given by $\lb \xi\xi \rb= \hbar {\bf N}(0)$ and if one uses a cut-off function of the form $f(\nu)=e^{-\nu/\Lambda}$ we obtain
\beq
{\bf N}=\frac{1}{2}\tilde{\gamma}^2 \int_0^{\infty} d\nu \nu e^{-\nu/\Lambda} =\frac{1}{2} \tilde{\gamma}^2 \Lambda^2.
\eeq

Therefore $\xi_{\rm rms}= \sqrt{\hbar/2} \tilde{\gamma}\Lambda$, and
\beq
\tilde{\gamma}= \gamma \frac{\sqrt{\rho}}{\sqrt{\hbar/2}\Lambda\tau}(v_{\text{max}}-v_{\text{min}}).
\eeq

If we assume that $\tau$ is the minimum amount of time during which the distribution changes appreciably (as we did to compute the mean value of the force) and it is of the same order that the magnitude of the time scale associated to the environment, then the Lorentzian cut-off frequency satisfy the relation $\Lambda \tau \approx 1$ and we finally arrive to the desired expression
\beq
\tilde{\gamma} = \gamma \frac{\sqrt{2\rho}}{\sqrt{\hbar}}(v_{\text{max}}-v_{\text{min}}).
\eeq

In accordance to the results presented in the Ref. \cite{cirac} the bound of the coupling constant proposed originally is $\gamma(\tilde{\gamma})\leq 5\cdot10^{-6}$. As explained in the previous section, this bound comes from the requirement that the fluctuations in the trajectories due to the noise in the force do not wash out the characteristic peak in the correlation.

 In the following section we will learn that this bound is not appropriate since in this range decoherence occurs even before the acoustic black hole is formed, at a time of the order of $\tau$. To increase the decoherence time to values bigger than the measurement time, of order $T$, the magnitude of the noise present in the system, characterized by $\gamma$, must be appropriately reduced.
\section{Decoherence Time}\label{sec:tdecc}
\subsection{ Master equation for the reduced density matrix.}
To obtain the decoherence time first one has to obtain the relevant coefficients included in the master equation. In particular, those corresponding to diffusive effects. In order to obtain this equation, the procedure consists of deriving the evolution operator with respect to time to generate the derivative of the reduced density matrix. The next step is to multiply $\mathcal{J}_r \rho_r(t_i)$ and finally integrate the initial conditions thus generating $\rho_r(t)$ instead of its initial value \cite{lombardo,calhu}. 
\begin{widetext}
The derivative of $\mathcal{J}_r$ is 
\begin{equation}
i\hbar \frac{\partial}{\partial t} \mathcal{J}_r[\phi_f,\phi'_f,t|\phi_i,\phi'_i,0] = \mathcal{J}_r[\phi_f,\phi'_f,t|\phi_i,\phi'_i,0] \bigg\{ -\frac{\partial}{\partial t}  S_{\text{eff}}[\phi_{\text{cl}},\phi'_{\text{cl}},t] \bigg\},
\end{equation}
and using the following identity that comes from the causality of the dissipation kernel
\begin{equation}
S_{\text{IF}}[\phi,\phi',t+dt]=S_{\text{IF}}[\phi,\phi',t]+dt \int dx \phi^-(x,t) \int_0^t ds \{ {\bf D}(t,s) \phi^+(x,s)+i{\bf N}(t,s)\phi^-(x,s)\},
\end{equation}
one can obtain, after performing the functional integrals
\bea
\frac{\partial}{\partial t}\rho_r(\phi,\phi';t)&=&-\frac{i}{\hbar} \lb \phi|[\hat{H}_{\mathcal{S}},\hat{\rho}_r(t)]|\phi'\rb- \frac{1}{\hbar}\int dx (\phi(x)-\phi'(x))\int_0^t dt'  \nonumber\\
&&\times\bigg\{ {\bf N}(t,t')[\Phi-\Phi'](\phi,\phi';t')- \frac{i}{2}{\bf D}(t,t')[\Phi+\Phi'](\phi,\phi',t)\bigg\}.
\ea
The first term that comes from deriving the free part of the effective action generates the usual Liouville-von Neumann term. The second contribution was defined as
\beq
\Phi(\phi,\phi';t')\equiv\int_{\phi(t,x)=\phi(x),\phi'(t,x)=\phi'(x)}\mathcal{D}x\mathcal{D}x'~e^{iS_{\rm eff}/\hbar}\rho_r(\phi_0,\phi'_0;0)\phi(x,t'),
\eeq
and a similar expression for $\Phi'[\phi,\phi';t']$ with $\phi(t')\mapsto \phi'(t')$. In the case of weakly coupled environments, $\gamma\sim10^{-6}$ and since this terms are already beyond linear order with respect to the coupling constant because of the presence of the kernels, the expression above can be approximated by $S_{\rm eff}[\phi,\phi']\simeq S_0[\phi]-S_0[\phi']$ in the functional integral. This way, we can see that this functions are the matrix elements of the operators $\Phi(x,t')=\hat{\phi}(x,t'-t)\rho_r(t)$ y $\Phi'(x,t')=\rho_r(t)\hat{\phi}(x,t'-t)$. Using this expression we can finally obtain the master equation from
\beq
\hbar\frac{\partial}{\partial t}\hat{\rho}_r(t)=-i[\hat{H}_s,\hat{\rho}_r(t)]-\int_0^td\tau dx\bigg\{ {\bf N}(t,t-\tau)[\hat{\phi}_S(x),[\hat{\phi}(x,-\tau),\hat{\rho}_r(t)]]-\frac{i}{2}{\bf D}(t,t-\tau)[\hat{\phi}_S(x),\{\hat{\phi}(x,-\tau),\hat{\rho}_r(t)\}]\bigg\}, \label{ema}
\eeq
where $\hat{\phi}_S(x)$ is the field operator in the Schr\"odinger picture and $\hat{\phi}(x,-\tau)$ corresponds to the Heisenberg one. To study the decoherence, we are only interested in the term proportional to the noise kernel since those are the ones that generates the necessary diffusion terms in the master equations. To achieve this, we need to replace the solution of the Heisenberg equations of motion (EOM).
\end{widetext}

To continue we will obtain the solutions of the Heisenberg EOM, which coincide with the classical EOM of the field. Since we are in $1+1$ dimensions, the solution can be cast in the following form
\beq
\phi=f({\sf u})+g({\sf v}),
\eeq
where we define the null coordinates associated with the effective metric
\begin{eqnarray*}
 {\sf u}&=& t-\int \frac{dx}{c(x)+v(x)}=t-x_{\sf u}\\
 {\sf v}&=& t+\int \frac{dx}{c(x)-v(x)}=t-x_{\sf v}.
\end{eqnarray*}
Therefore, we can study the decoherence for each mode ${\sf u}$ and each mode ${\bf v}$ separately
\beq
\hat{\phi}(x,t)=\hat{\phi}_{{\sf u}/{\sf v}}\cos{\omega(t-x_{{\sf u}/{\sf v}})}.
\eeq
The allowed frequencies $\omega$ can be found requiring the following condition since the spatial dimension is compact, $x\equiv x+2n\pi$, and one has to impose periodic boundary conditions over the ring $\cos{\omega(t-x_{{\sf u}/{\sf v}}(x+2n\pi))}=\cos{\omega(t-x_{{\sf u}/{\sf v}}(x))}$. We take as the maximum allowed frequency to be $\omega_{\rm max}\sim N/T$. The solution is continuous in the subsonic-supersonic transition regions. The contribution proportional to the canonically conjugate momentum $\hat{\Pi}_S$ can be discarded since it does not generate a diffusive term.

We will study separately the decoherence present in each mode of frequency $\omega$ and both ${\sf u}$ and ${\sf v}$ modes. Inserting this solution of the EOM in Eq. (\ref{ema}) the relevant term is given by
\bea
\int_0^td\tau dx &&{\bf N}(t,t-\tau)[\hat{\phi}_{{\sf u}/{\sf v}}\cos{\omega x_{{\sf u}/{\sf v}}},\nonumber\\
&&[\hat{\phi}_{{\sf u}/{\sf v}}\cos{\omega(\tau+x_{{\sf u}/{\sf v}})},\hat{\rho}_r(t)]].
\ea
Taking the appropriate expectation value one obtains
\bea
&&(\phi_{{\sf u}/{\sf v}}-\phi'_{{\sf u}/{\sf v}})^2\int_0^td\tau dx {\bf N}(t,t-\tau) \cos{\omega x_{{\sf u}/{\sf v}}}\nonumber\\
&&\times\cos{\omega (x_{{\sf u}/{\sf v}}+\tau)}\rho_r(\phi,\phi')\nonumber\\
&=&(\phi_{{\sf u}/{\sf v}}-\phi'_{{\sf u}/{\sf v}})^2\int_0^td\tau dx {\bf N}(t,t-\tau) \cos{\omega x_{{\sf u}/{\sf v}}}\nonumber\\
&&~\bigg(\cos{\omega x_{{\sf u}/{\sf v}}}\cos{\omega\tau}-\sin{\omega x_{{\sf u}/{\sf v}}}\sin{\omega\tau}\bigg)\rho_r(\phi,\phi').\nonumber
\ea
To estimate the decoherence time we study trajectories with $(\phi_{{\sf u}/{\sf v}} -\phi'_{{\sf u}/{\sf v}})^2\sim \rho \delta^2$, where $\delta=2\pi/N$, the mean separation of the ions since it is the smallest length scale of the system. We use the usual definition of the normal and anomalous diffusion coefficients respectively
\bea
D(t)&=&\int_0^td\tau{\bf N}(t,t-\tau)\cos{\omega\tau}\nonumber\\
f(t)&=&\int_0^td\tau{\bf N}(t,t-\tau)\sin{\omega\tau},
\ea
and we also define the following coefficients
\bea
V_{1{\sf u}/{\sf v}}&=&\int_0^{2\pi} dx(\cos{\omega x_{{\sf u}/{\sf v}}})^2\nonumber\\
V_{2{\sf u}/{\sf v}}&=&\int_0^{2\pi} dx\cos{\omega x_{{\sf u}/{\sf v}}}\sin{\omega x_{{\sf u}/{\sf v}}}.
\ea
Therefore, the diffusion term we obtain is given by
\beq
\hbar\dot{\rho}_r \sim - \rho\delta^2 \{ D(t) V_{1{\sf u}/{\sf v}}+f(t)V_{2{\sf u}/{\sf v}}\}\rho_r.
\eeq
This term contributes in the following way to the solution of the master equation
\beq
\rho_r \sim \rho_r^U \text{exp}\bigg\{ -\frac{1}{\hbar} \rho \int_0^t dt'(D(t')V_{1{\sf u}/{\sf v}}+f(t')V_{2{\sf u}/{\sf v}}) \delta^2\bigg\},
\eeq
where $\rho^U$ represents the unitary evolution of the reduced density matrix. The decoherence time is defined based on the following relation 
\beq
\frac{ \rho}{\hbar} \int_0^{t_D} dt'(D(t')V_{1u/v}+f(t')V_{2u/v}) \delta^2 \approx 1.
 \eeq
 This corresponds to the fact that for times $t>t_D$ the non diagonal elements of the density matrix that encompass the quantum coherence effects are suppressed. In order to compute it, we will need explicit expressions for $D(t)$ y $f(t)$. 

Lets start with the normal diffusion coefficients at zero temperature $T_0=0$
\beq
D(t)=\frac{\tilde{\gamma}^2}{2} \int_0^{\infty} d\nu \int_0^t ds \frac{\nu}{1+\big(\frac{\nu}{\Lambda}\big)^2} \cos{\nu s} \cos{\omega s},
\eeq
where the expression was written with a Lorentzian cut-off to attenuate high frequencies. According to Ref. \cite{paula}, the explicit calculation of this integral gives the following result
\begin{eqnarray*}
D(t)=&&\frac{\tilde{\gamma}^2}{2} \frac{\omega}{1+\big(\frac{\omega}{\Lambda}\big)^2}\bigg[ \text{Shi}(\Lambda t)\bigg( \frac{\Lambda}{\omega}\cos{\omega t}\cosh{\Lambda t}\nonumber\\
&&+\sin{\omega t}\sinh{\Lambda t}\bigg)-\text{Chi}(\Lambda t)\bigg( \frac{\Lambda}{\omega}\cos{\omega t}\sinh{\Lambda t}\nonumber\\
&&+\sin{\omega t}\cosh{\Lambda t}\bigg)+\text{Si}(\omega t) \bigg],
\end{eqnarray*}
where $\text{Shi}(x)$ and $\text{Chi}(x)$ are the hyperbolic sine and cosine integral respectively and $\text{Si}(x)$ is the sine integral.

On one hand, for times much larger than the frequency cut-off scale, i.e. $t\gg \Lambda^{-1}$, this expression can be approximated as
\begin{equation}
D(t)\approx\frac{\tilde{\gamma}^2}{2} \frac{\omega}{1+\big(\frac{\omega}{\Lambda}\big)^2}\text{Si}(\omega t).
\end{equation}
On the other hand, for times much larger than the scale imposed by the field frequency, $t\gg \omega^{-1}$, the sine integral can be approximated as $\text{Si}(\omega t)\approx\pi/2$ and for frequencies much lower than the cut-off $\omega \ll \Lambda$,  $1/1+\big(\frac{\omega}{\Lambda}\big)^2\approx 1$. Finally, one obtains the expression
\beq
D(t)\approx \tilde{\gamma}^2\omega \frac{\pi}{4}.
\eeq
Therefore, the integral of the coefficient is given by
\beq
\int_0^{t_D}dsD(s)\approx \tilde{\gamma}^2\frac{\pi}{4} \omega t_D.
\eeq

The same analysis can be done for the anomalous diffusion coefficient and can be found in \cite{paula}. The result for $t\gg\omega^{-1},\Lambda^{-1}$ is
\beq
f(t)\approx \frac{\tilde{\gamma}^2}{2}\pi \omega \log{\frac{\Lambda}{\omega}}.
\eeq
As in the previous section, we can estimate the characteristic time of the environment as $\Lambda^{-1}\sim \tau$, then after performing the integral, the anomalous diffusion coefficient can be estimated as
\beq
\int_0^{t_D}dsf(s)\approx \frac{\tilde{\gamma}^2}{2} \pi \log{\frac{1}{\omega\tau}}\omega t_D.
\eeq
\begin{figure}[h]
\begin{center}
\subfigure[]{
\includegraphics[width=0.46\textwidth,height=0.22\textheight]{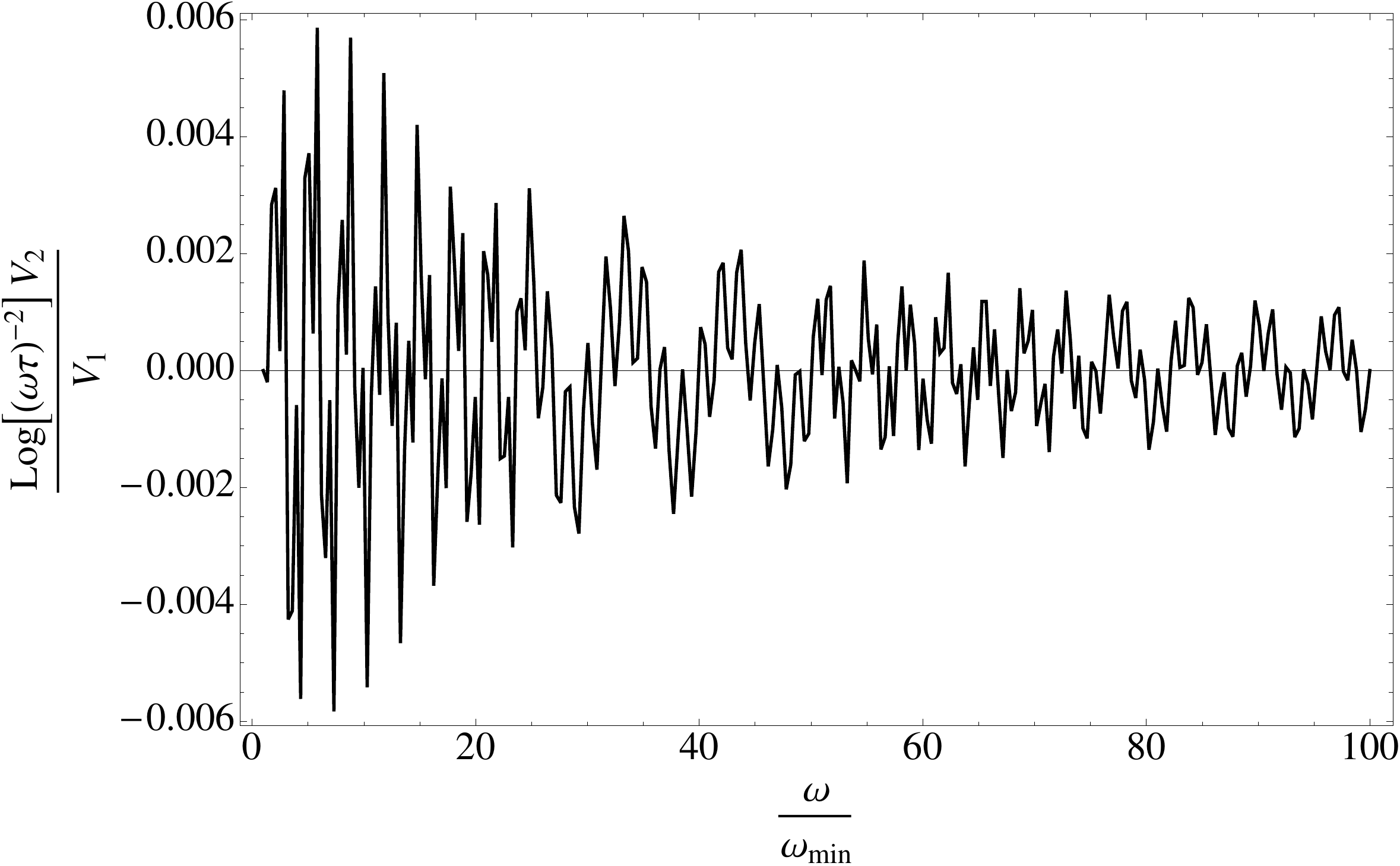}}
\subfigure[]{
\includegraphics[width=0.45\textwidth,height=0.22\textheight]{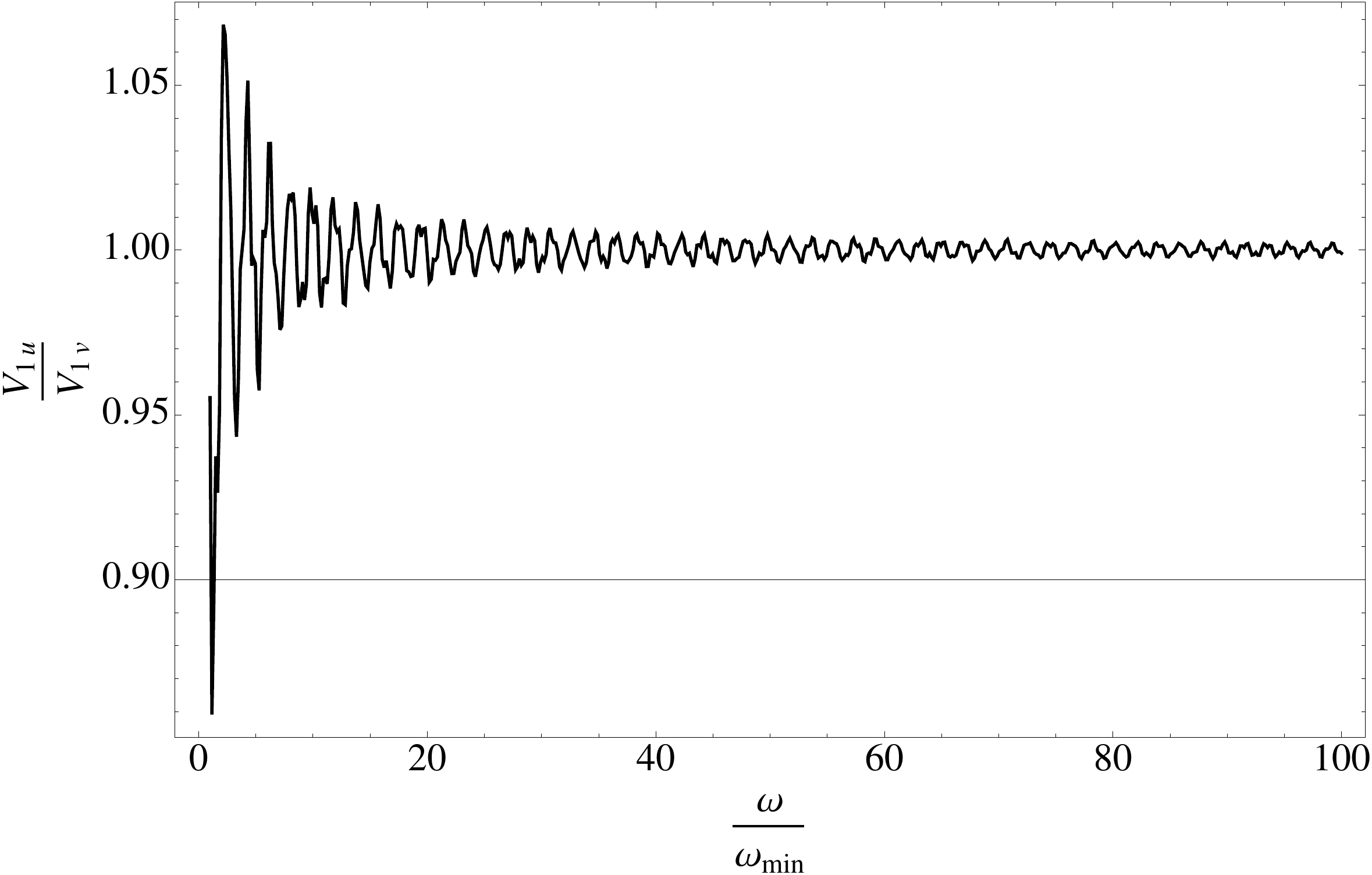}}
\end{center}
\caption{\label{fig:v} Plot of the functions (a) $\log{(\omega\tau)^{-2}}V_{2{\sf u}/{\sf v}}\ll V_{1{\sf u}/{\sf v}}$, showing that the anomalous component is negligible; and (b) $V_{1{\sf u}}/V_{2{\sf v}}$ with respect to the mode frequency $\omega$, showing that the decoherence time is the same for both ${\sf u}$ and ${\sf v}$ modes.}
\end{figure}
Putting the previous results together, for an initial state with zero temperature $T_0=0$, the decoherence time is given by the following expression
\bea
 t_D(T_0=0) &\approx& \frac{4\hbar}{\rho \delta^2 \tilde{\gamma}^2 \pi \omega(V_{1{\sf u}/{\sf v}}+2\log{\frac{1}{\omega\tau}} V_{2{\sf u}/{\sf v}}) }\nonumber\\
 &=& \frac{2\hbar^2}{\gamma^2 \Delta v \delta^2 \omega \pi  \rho^2}\nonumber\\
 &&~~\times (V_{1{\sf u}/{\sf v}}+2\log{\frac{1}{\omega\tau}} V_{2{\sf u}/{\sf v}})^{-1},
 \ea
where $\Delta v\equiv v_{\rm max}-v_{\rm min}$. If we focus in the ions' ring and the experimental parameters proposed for the system, then the expression can be simplified even more, since we are in a possition to approximate
\beq
2\log{\bigg(\frac{1}{\omega\tau}\bigg)}V_{2{\sf u}/{\sf v}}\ll V_{1{\sf u}/{\sf v}}~~~\text{and}~~~V_{1{\sf u}}\simeq V_{2{\sf v}}\equiv V,
\eeq
as can be verified in the plots shown in Fig. \ref{fig:v}. Finally, we obtain the following result for the decoherence time
\beq
\label{tdecsint}
t_D(T_0=0)=\frac{2\hbar^2}{\gamma^2 \Delta v\delta^2\omega\pi\rho^2V}.
\eeq
\begin{figure}[ht]
\begin{center}
\includegraphics[width=0.48\textwidth]{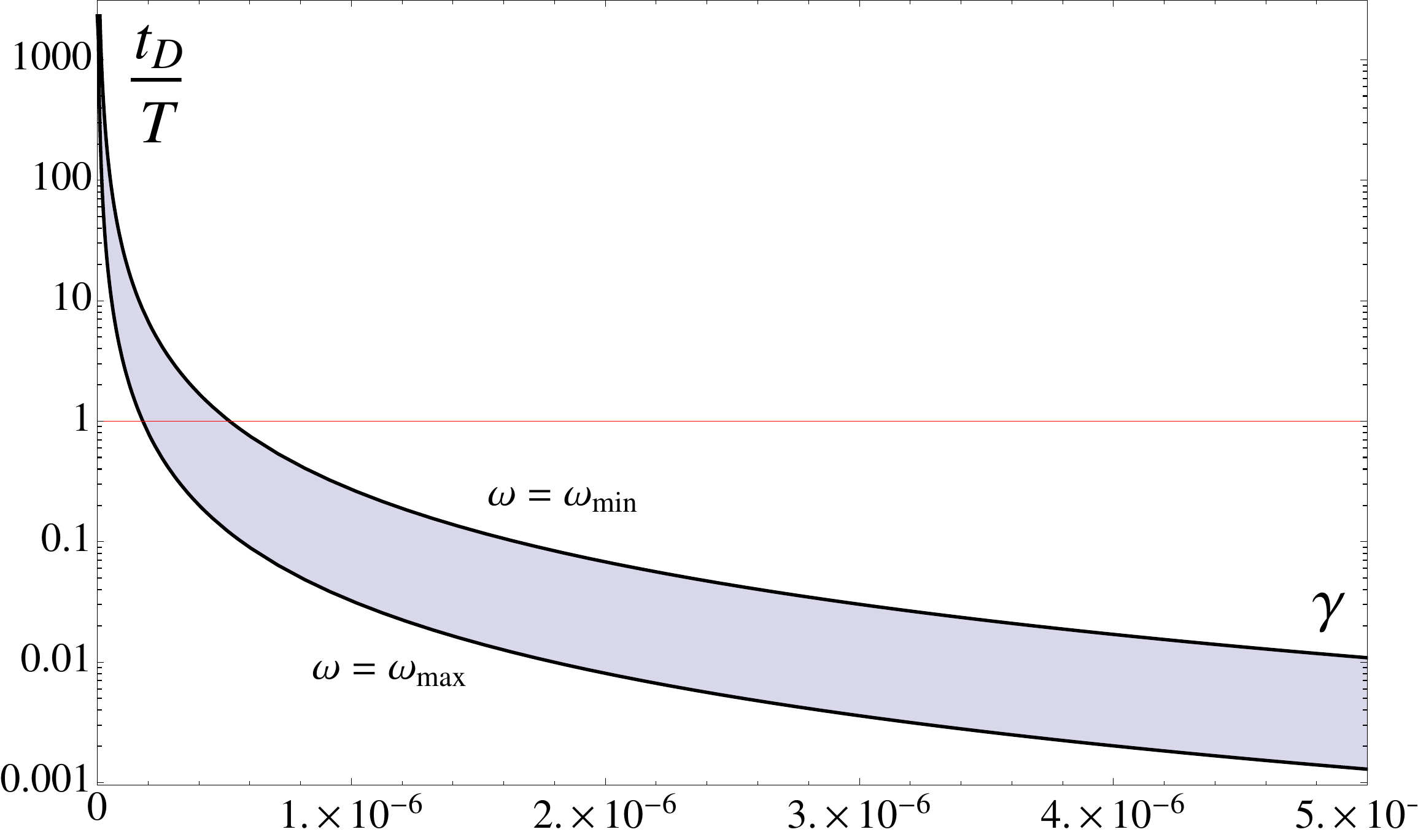}
\end{center}
\caption{\label{tdec1}[Color Online] Dependence of the decoherence time as a function of the relative noise in the force $\gamma$. The colored band includes all the allowed frequency range of the solutions.}
\end{figure}
In Fig. \ref{tdec1} the result for the decoherence time as a function of the force relative noise $\gamma$ is shown. The decoherence time is much shorter that $T$ for the bound presented in the Ref. \cite{cirac}, $\gamma\sim5\cdot10^{-6}$. As explained in section \ref{sec:modelo}, the period $T$ is also of the same order of magnitude as the time needed to perform the measurement. For times larger, the system becomes classically unstable and for times smaller the black hole would have no time to develop. The system, for the experimental parameters proposed in Ref. \cite{cirac}, shows decoherence in a time-scale too short and this would make impossible the measurement of the aspects of the Hawking effect, which is purely of a quantum nature. Moreover, decoherence happens in a time of the same order of magnitude as the collapse $t_D\sim\tau$, therefore there is no time for the acoustic black hole to form.

To find a solution for this issue, we impose a decoherence time appropriate to the experiment, to be specific, no shorter than the following bound $t_D\geq 100\cdot T$.

For this to be satisfied and to avoid the quantum to classical transition induced by the environment, the bound for the coupling with the environment must be modified in the following way
\beq
\gamma \lesssim 3\cdot10^{-8}.
\eeq
\begin{figure}[ht]
\begin{center}
\includegraphics[width=0.48\textwidth]{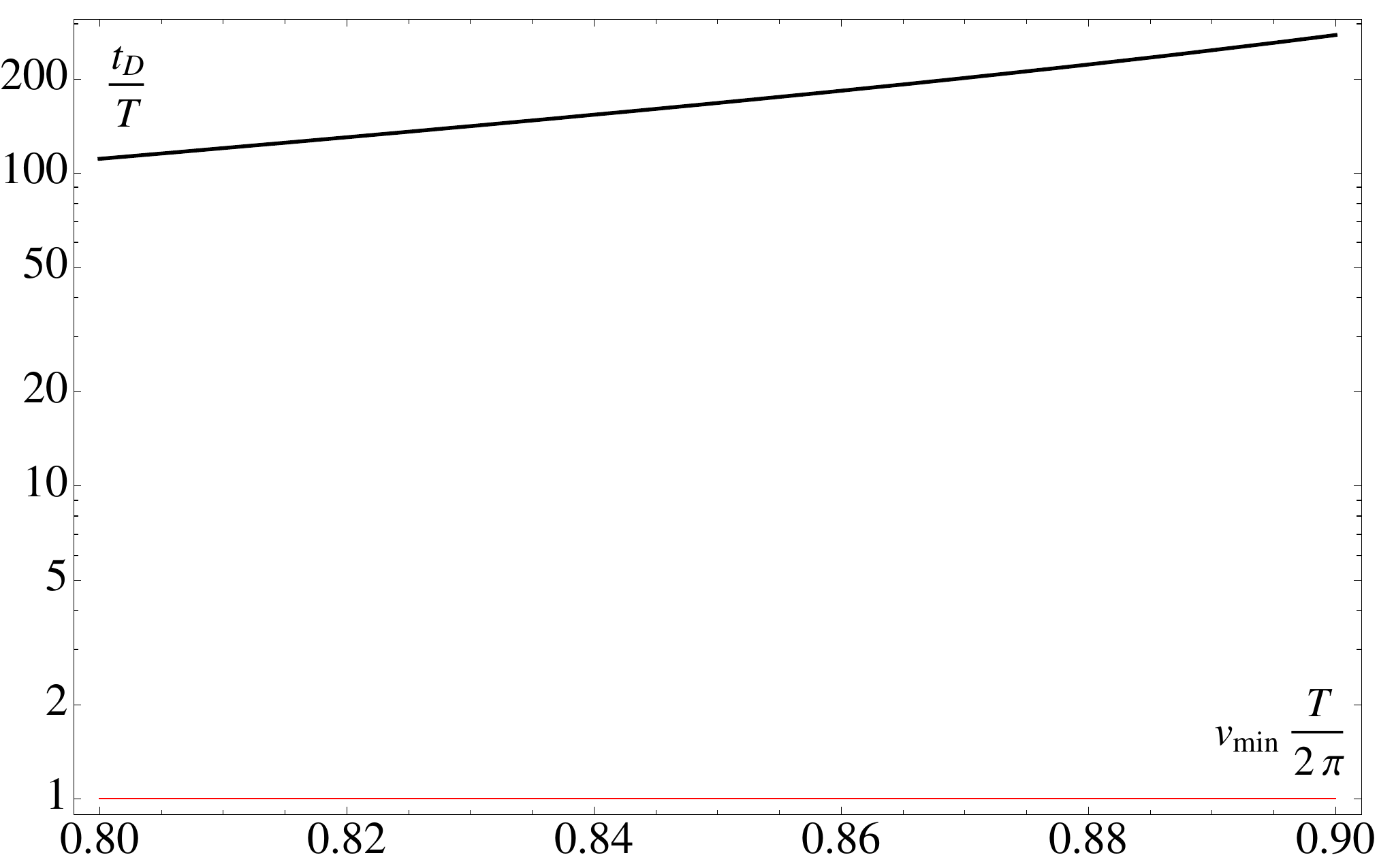}
\end{center}
\caption{\label{tdec5}[Color Online] Dependence of the decoherence time as a function of the minimum velocity $v_{\rm min}$.}
\end{figure}
We also study the dependence of the decoherence time with the shape of the velocity profile. To do that we plot the expression in Eq. (\ref{tdecsint}) as function of $v_{\rm min}$. As explained in Ref. \cite{cirac} this magnitude must be near $0.8\overline{3}2\pi/T$ to perform the measurement. As observed in Fig. \ref{tdec5}, the decoherence time varies smoothly with the velocity, therefore small deviations over $0.8\overline{3}\cdot2\pi/T$ do not present inconvenience with respect to decoherence as long as $\gamma\leq10^{-8}$ and $T_0\lesssim100\cdot T_H$.
\subsection{Behavior of the Decoherence Time with Temperature.} \label{temtdec}
In this section we will study how temperature influences the previous results. To achieve this we need to calculate the diffusion coefficients again, in the presence of nonzero temperature
\beq
D(t)=\frac{\tilde{\gamma}^2}{2} \int_0^{\infty} d\nu \int_0^t ds \nu \coth{\frac{\beta\hbar\nu}{2}} \cos{\nu s} \cos{\omega s}.
\eeq
To calculate this integral in the low temperature limit we will use the following approximation
\beq
\coth{\frac{\beta\hbar\nu}{2}} \approx \begin{cases}
\frac{2}{\beta\hbar\nu}    &  \text{si }\frac{\beta\hbar\nu}{2}\ll1\\
1   & \text{si }\frac{\beta\hbar\nu}{2}\gg1.
\end{cases}
\eeq
\begin{figure}[h!]
\begin{center}
\includegraphics[width=0.48\textwidth]{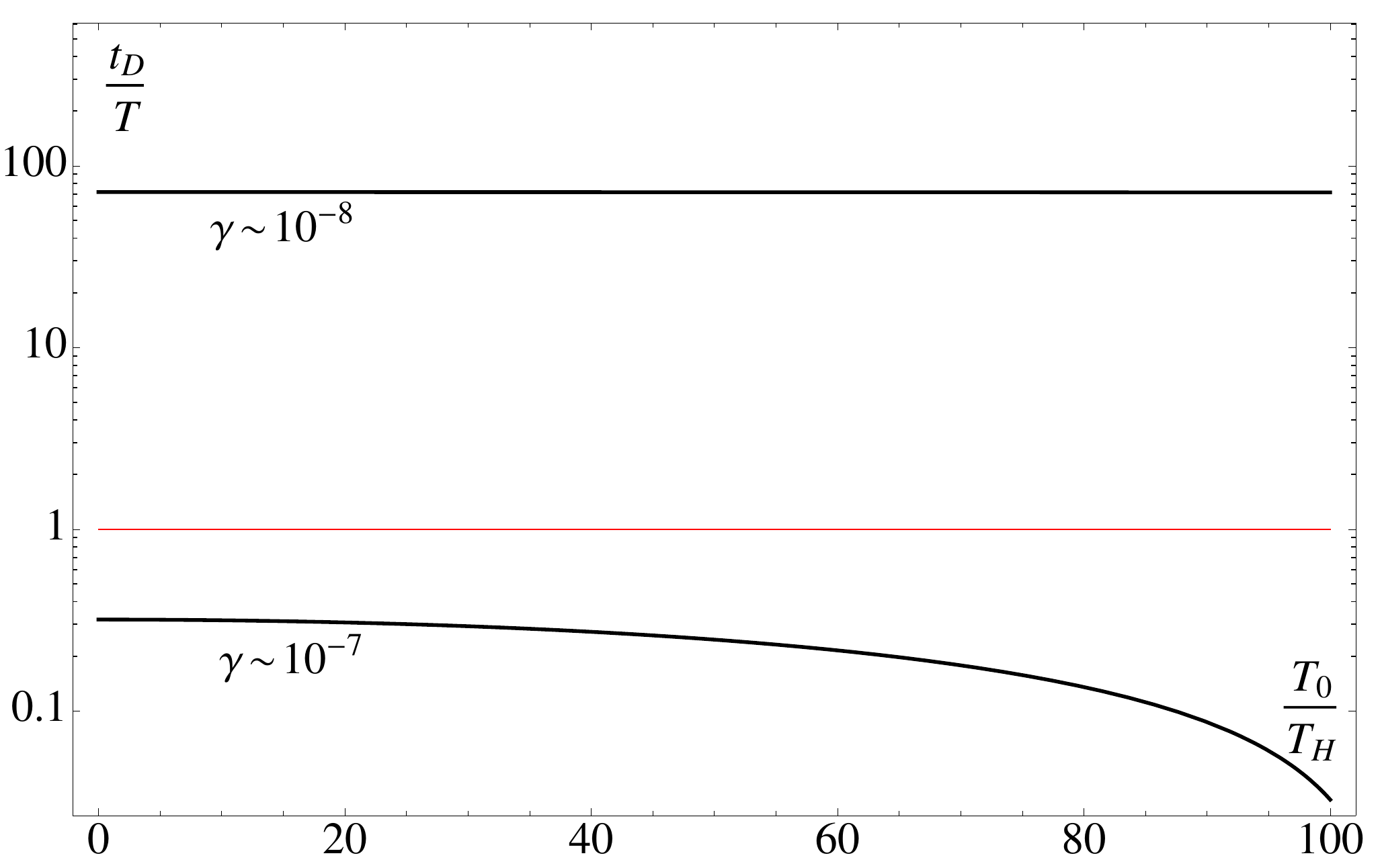}
\end{center}
\caption{\label{tdec2}[Color Online] Dependence of the decoherence time for different relative noises in the force as a function of temperature.}
\end{figure}
and perform the integrals in the two regimes, both $\frac{\beta\hbar\nu}{2}\ll1$ and $\frac{\beta\hbar\nu}{2}\gg1$. The details of the calculation can be found in appendix \ref{A}. The result for the integral of the diffusion coefficient to lowest order in temperature is given by
\bea
\int_0^{t_D} dt D(t,\beta)&=&\tilde{\gamma}^{2}\frac{\omega \pi}{4} t_D+\tilde{\gamma}^{2} \frac{2(1- \cos{\omega t_D})}{2\omega^2\beta^2\hbar^2}\nonumber\\
&&\lesssim \tilde{\gamma}^{2}\frac{\omega \pi}{4} t_D+ \frac{2\tilde{\gamma}^{2}}{\omega^2\beta^2\hbar^2}.
\ea
Using this expansion we can conclude that the decoherence time at finite temperatures is
\beq
\label{tdec}
t_D(T_0)=\frac{2\hbar^2}{\gamma^2 \Delta v\delta^2\pi\omega \rho^2V}-\frac{8k_B^2T_0^2}{\omega^3\pi\hbar^2}+\mathcal{O}(T_0^3).
\eeq

Taking advantage of the expression in Eq. (\ref{tdec}), lets also study the dependence of the decoherence time with the initial temperature of the field $T_0$ \footnote{In general, $T_0$ is the initial temperature of the environment. Since we consider that initially the system is in thermal equiibrium this is also the initial temperature of the system.}. The result is plotted in the Fig. \ref{tdec2}. One can observe in the figure that the decoherence time does not change substantially with temperature for small enough coupling and therefore this makes no restriction on the temperature. Decoherence aside, taking into account the results obtained in Ref. \cite{cirac}, the entanglement that one would wish to measure as a signature for the Hawking effect is present only for $T_0\lesssim 100\cdot T_H$. It is also important to notice that the low temperature approximation is valid in the range $T\lesssim 100T_H$.

 Here we assumed that the system is initially in thermal equilibrium with respect to the environment. However, if the bath is at a higher temperature than the system one would expect that the bath heats up the system and so washes out the Hawking signal. This featured is reflected in the fact that the decoherence time decreases for increases temperatures $T_0$, which must now be interpreted as the initial temperature of the bath. 
\section{Dynamics of the Entanglement through Correlations} \label{sec:ep}
In this section we will study the signature of the Hawking effect through the shape of the correlation for a generic acoustic black hole, following the presentation given in Ref. \cite{tesis}.

Hawking radiation can be understood as a pair production of virtual particles, one of which falls into the black hole and the other, outgoing, becomes real, building up the Hawking radiation \cite{hawking}, as can be seen from Eq. (\ref{squeezed}). The role of the correlations between this pair of particles has been studied as a signature of the quantum Hawking effect, see for example Refs. \cite{corrbec,entre:unruh-cor,cirac}. It was found that the entanglement between this pair is translated to a sharp peak of $\langle \Pi_L(x_1,t)\Pi_L(x,t) \rangle$ as a function of $x$, where $x_1$ is inside the black hole and $x$ is outside, $\Pi$ is the canonical momentum conjugate to $\phi$; the subscript $L$ corresponds to the left-moving modes. This magnitude was calculated using the Israel-Hartle-Hawking state \cite{Boulware,entre:birrel-davies,IHH,entre:unruh-cor} and it was also calculated numerically in the case of the circular ion trap \cite{cirac} and BEC \cite{corrbec}. The result is shown in \cite{entre:unruh-cor}, Fig. 1.

The formalism used in Ref. \cite{entre:unruh-cor} does not seem to be useful in the case of an acoustic black hole as an open quantum system. The presence of the environment does not seem to be easily included. Moreover, in the computation a model of \emph{eternal} black hole is used. Therefore, the inclusion of an initial thermal state and the time evolution of the correlation function cannot be obtained within previous procedures.

In this section we want to resolve this issues to some extent. We will analyze the complete evolution starting from the flat background coupled to a massless scalar particle, initially in a thermal state of temperature $T_0$, and ``collapsing" toward an acoustic black hole background with an event horizon.

To find the correlation functions in the quantum regime, we use the stochastic formalism developed by Calzetta \emph{et al} in Ref. \cite{calzetta}, reviewed in Ref. \cite{calhu} and applied in Ref. \cite{tesis} to this problem. This procedure will allow us to find the correlation functions through the semiclassical solutions of the Langevin EOM. We will assume that we study a generic acoustic black hole with a generic environment. The coarse grained effective action of a generic field, that emerges for example from the presence of an environment, can be cast into the form
\beq
S_{\rm eff}[\phi,\phi']=S_0[\phi]-S_0[\phi']+S_{IF}[\phi,\phi'],
\eeq
where the free action is given by Eq. (\ref{ac}), the influence action by Eq. (\ref{if}) and the kernels ${\bf D}(x,x')$ and ${\bf N}(x,x')$ include all the relevant information regarding the environment. In the case of harmonic oscillators with a spectral density their expression was already given in Eqs. (\ref{d}) and (\ref{n}).
\subsection{Stochastic Description}
The purely imaginary term of the effective action can be written in the following way\footnote{In this section, to simplify the calculation we use $c=\hbar=1$.}
\beq
\int\mathcal{D}\xi~P[\xi]e^{-i \phi^-_x\xi_x}=e^{i \frac{i}{2}\phi^-_x{\bf N}_{x,x'}\phi^-_{x'}},
\eeq
where, for example $\phi_x=\phi(x)$ and repeated indexes are integrated. The probability distribution is gaussian
\beq
P[\xi]\equiv N_{\xi} e^{ -\frac{1}{2} \xi_x {\bf N}^{-1}_{x,x'}\xi_{x'}},
\eeq
where $N_{\xi}$ is a normalization constant, see Ref. \cite{lombardo,calhu,calzetta}. Therefore, the original action is the same as a stochastic process described by an action of the following type
\beq
S[\phi,\phi']=S_0[\phi]-S_0[\phi']+\phi^-_x{\bf D}_{x,x'}\phi^+_{x'}-\phi^-_x\xi_x.
\eeq
We will start with general results regarding the correlation function and later use it to the particular case of acoustic black holes.

To find the correlation functions we must solve the semiclassical EOMs for a realization of the stochastic force $\xi(t,x)$ and then integrate over the possible solutions with a weight imposed by the probability distribution $P[\xi]$ and also integrate over the initial conditions. It is in the latter integration that the initial temperature of the field $T_0$ makes its appearence.

To start, we will make a distinction between the homogeneous and inhomogeneous part of the solution, i.e.
\beq
\phi_{\xi}(x,t)=\phi_O(x,t)+\int_0^t dsdx' G_{\text{ret}}(t,s|x,x') \xi(s,x')
\eeq
where $\phi_O(x,t)$ is the noiseless solution satisfying
\beq
 \Box\phi_O(t,x)+\int_0^tds{\bf D}(t,s)\phi_O(s,x)=0, \label{eq:lange}
\eeq
where $\Box\equiv \nabla_{\mu}\nabla^{\mu}$ and $G_{\text{ret}}$ is the retarded Green function associated with the equation, satisfying
\beq
\Box G_{\text{ret}}+\int_0^tdt'{\bf D}(t,t')G_{\text{ret}}(t',s)=\delta(t-s)\delta(x'-x).
\eeq
We take the usual Painlev{\'e}-Gullstrand-Lema{\^\i}tre metric associated to a generic acoustic black hole
\bea\label{entre:metrica}
ds^2&=&-dt^2+(dx-v(x,t)dt)^2 \nonumber\\
&=&-(1-v(x,t)^2)dt^2-2v(x,t)dxdt+dx^2, 
\ea
where $v(x,t)$ is the velocity profile, in principle arbitrary as long as it has a supersonic subsonic transition, i.e. an event horizon. In this section we now consider its time dependence instead.

In this paper we work under the weak coupling approximation, in such a way that we can estimate $\phi_O$ beginning with the zeroth order limit of the EOM $ \Box \phi_O\simeq0$ which explicitly, replacing the metric of Eq. (\ref{entre:metrica}), is given by
\beq\label{eq:aproxcoup}
[(\partial_t +\partial_x v(x,t))(\partial_t+v(x,t)\partial_x)-\partial^2_x]\phi_O(x,t)=0,
\eeq
and moreover, we can approximate $G_{\text{ret}}$ from
\begin{equation}
\Box G_{\text{ret}}(t,s|x,x')\simeq\delta(t-s)\delta(x'-x).
\end{equation}
Therefore to first order in the coupling with the environment we can consider the dissipation term in Eq. (\ref{eq:lange}) as an inhomogeneity, replacing the complete solution $\phi_O$ by the solution of the Eq. (\ref{eq:aproxcoup}), 
\beq
\Box \phi\simeq\xi(x,t)-\int_0^tdt'{\bf D}(t,t')\phi_O(x,t'),
\eeq
which in turn has the solution
\begin{equation}
\phi_{\xi}(x,t)\simeq\phi_O(x,t)+\int_0^t dsdx' G_{\text{ret}}(t,s|x,x')\xi_f(s,x'),
\end{equation}
where we define $\xi^f$ as
\beq
\xi^f(t,x)=\xi(t,x)-\int_0^tds{\bf D}(t,s)\phi_O(x,s).
\eeq
Assuming that we have this solutions, the correlation function relevant to the study of the entanglement between the Hawking pair phonons can be obtained from 
\beq
\lb \phi_1\phi_2 \rb \equiv \lb\lb\phi(x_1,t_1)\phi(x_2,t_2) \rb_{\xi} \rb_{\text{in}},
\eeq
where $\lb \ldots \rb_{\text{in}}$ represents the mean value integrated over the initial conditions weighted with the thermal distribution with $T_0$. To obtain the correlation function between the conjugate momentums one simply has to derive the equation above following the definition below
\beq
\Pi(x,t)=\frac{\partial \phi}{\partial t}+v(x,t)\frac{\partial \phi}{\partial x}. \label{momento}
\eeq 
Therefore, following the formalism presented in \cite{calzetta},
\bea
\langle \phi_1\phi_2 \rangle&=& \lb (\phi_{O1}+G_{1x}\xi^f_x)(\phi_{O2}+G_{2x}\xi^f_x) \rb_{\xi, {\rm in}} \nonumber\\
&=& \langle \phi_{O1}\phi_{O2} \rangle_{\text{in}}- G_{2x'}{\bf D}_{x'x}\langle\phi_{Ox}\phi_{O2}\rangle_{\rm in}\nonumber\\
&&- G_{2x'}{\bf D}_{x'x}\langle\phi_{Ox}\phi_{O1}\rangle_{\rm in}\nonumber\\
&&+G_{1x'}\lb \xi_{x'}\xi_{x}\rb_{\xi,{\rm in}}G_{2x}.
\ea
Using the properties of the stochastic force probability distribution we get
\bea
\langle \phi_1\phi_2 \rangle&&=\langle \phi_{O1}\phi_{O2} \rangle_{\text{in}}- G_{2x'}{\bf D}_{x'x}\langle\phi_{Ox}\phi_{O2}\rangle_{\rm in}\nonumber\\
&&- G_{2x'}{\bf D}_{x'x}\langle\phi_{Ox}\phi_{O1}\rangle_{\rm in}+G_{1x'}{\bf N}_{x'x}G_{2x}.
\ea

In the case without environment, where the field $\phi$ represents particles in an acoustic black hole background, the previous formalism can be used but only the first term contributes. Therefore, in the general case one can write
\beq
\lb \phi_1\phi_2 \rb_O= \lb \phi_1\phi_2 \rb_C+\mathcal{O}(\lambda),
\eeq
where $\lambda$ is the coupling constant with the environment. The subindices $C$ and $O$ means that it corresponds to the closed and open system, respectively. This simple case dropping the $\mathcal{O}(\lambda)$ terms already presents the main difficulties inherent with the calculation and we will develop a technique to deal with the equations in the following sections.
\subsection{Modes for the Wave Equation.}
To solve the EOM for the field we will use the method of characteristics, well known from mechanics of compressible fluids, see Ref. \cite{fuidos}. If one has a general differential equation of the form 
\beq
a(x,t) \frac{\partial u}{\partial x}+b(x,t) \frac{\partial u}{\partial t}+c(x,t)u=0 ,\label{ec:car}
\eeq
with initial condition $u(x,0)=f(x)$ then it can be solved in the following way. First find the characteristic curves, defined as
\beq
\frac{dx}{ds}=a(x,t) \;\;\;\;\;\;\ {\rm and} \;\;\;\;\;\;\;\;\; \frac{dt}{ds}=b(x,t). \label{s}
\eeq
Then find the evolution of $u(x,t)$ along the characteristic curves finding the solution for the following ordinary differential equation
\beq
\frac{du}{ds}(x(s),t(s))+c(x(s),t(s))u(x(s),t(s))=0. \label{ode}
\eeq
Finally, having the congruence of characteristic curves, to know $u(x,t)$ invert $(x,t)\mapsto(x_0,s)$ where $x_0$ is the initial condition and $s$ the parameter along the characteristic that goes through $(x,t)$. Then, $u(x,t)=f(x_0)$.

In the rest of this section we will use this method to find the solution of the EOM for the field.

Before attempting to solve for the modes we have to give a specific velocity profile,
\beq
v(x,t)=\sigma(t)\times\begin{cases}
v_{\rm min} & {\rm para~} -\infty<x<-a\\
1+\kappa x & {\rm para~} -a<x<a \\
v_{\rm max} & {\rm para~} a<x<\infty,
\end{cases}
\eeq
where $\sigma(t)$ is a function that must satisfy $\sigma(0)=0$, to guaranty that at $t=0$ the metric is flat and $\sigma(t\gg\tau)=1$, in order for the background to ``collapse" in an stable acoustic black hole after the time interval $\tau$. A particular function that satisfies this requirement and is easy to manipulate is 
\beq
\sigma(t)=\tanh{\frac{t}{\tau}}.
\eeq
We will solve the equations first for the region $-a<x<a$ without taking into account the rest of the space. These solutions are the same in the other regions upon the following changes $\kappa=0$ and $\sigma\mapsto  v_{\rm max/min}\sigma$. After this, we will see how to introduce them properly and thus obtain the full solution. 

Nevertheless, the problem that the presence of different regions introduce are the interfaces. For example, a characteristic curve that starts in $-a<x_0<a$ eventually reach the point $x(s_i)=a$ and for $s>s_i$ the characteristic to use is the one corresponding to the region $a<x<\infty$. In this section we will see that the solutions that do not go through this boundaries do not present the characteristic peak associated with the entanglement of the Hawking pair. In the following section we will study the full solution and we will learn how the entanglement is developed and its relationship to this issues.

The equation we have to solve is Eq. (\ref{eq:aproxcoup}) which can be cast into the following form
\beq
(\partial_t +\partial_x v(x,t)+\partial_x)(\partial_t+v(x,t)\partial_x-\partial_x)\phi_O(x,t)=0.
\eeq
Defining the operators
\bea
\partial_L&=&\frac{\partial}{\partial t}+v(x,t)\frac{\partial}{\partial x}-\frac{\partial}{\partial x} \\
\partial_R&=&\frac{\partial}{\partial t} +\frac{\partial}{\partial x} v(x,t)+\frac{\partial}{\partial x}.
\ea
the equation can be written as $\partial_R\partial_L\phi=0$. One can first solve $\partial_L\phi_L=0$.  Then one has to solve $\partial_R\tilde{\phi}_R=0$ and finally $\partial_L\phi_R=\tilde{\phi}_R$ in order to obtain $\phi_R$. This way, the more general solution is $\phi_O=\phi_L+\phi_R$. Both left and right moving components can not be solved separately since $[\partial_L,\partial_R]\neq 0$.

The procedure to find the solution with the characteristic curves is developed in appendix \ref{solcar}. For example, the characteristic curves for the left moving modes are given by
\bea
x(t)=&&\exp{\bigg(\kappa\int_0^t\sigma(s)ds\bigg)}\nonumber\\
&&\times\bigg(x_0- \int_0^t (1-\sigma(t))e^{-\kappa\int_0^s\sigma(s')ds'} ds\bigg).
\ea
To make contact with the usual expansion in flat field theory we cast the solution into the following form
\beq
\phi(x)=\int dk (u_k(x)a_k + {\rm H.c.}), \label{entre:solmodos}
\eeq
where H.c. means taking the hermitian conjugate of the expression and the modes are given by
\bea
u_k(x)&=&\frac{1}{\sqrt{2|k|}} e^{ikxe^{-\kappa\int_0^t\sigma(s)ds}+ik \int_0^t (1-\sigma(s))e^{-\kappa\int_0^s\sigma(s')ds'} ds} \nonumber\\
&&\times\bigg\{1-\Theta(k)2i|k|\int_0^tds\; \nonumber\\
&&\times e^{-2ik \int_0^{s}e^{-\kappa\int_0^{s'}\sigma(s'')ds''} ds'-\kappa\int_0^s \sigma(s')ds'}\bigg\}.
\ea

The coefficients $a_k$ and $a^{\dagger}_k$ corresponds at early times to the creation and annihilation operators upon quantization and they would carry the subindex ``in".

To compute the correlation function when an environment is added we need the retarded Green function that we compute as $G_{\rm ret}(t,t'|x,x')=[\phi(x,t),\phi(x',t')] \Theta(t-t')$.
\subsection{Entanglement: Closed System.}
In this section we study the behavior of the two-point function of the left moving part of the momentum, $\Pi_L$. Using the expansion given in Eq. (\ref{entre:solmodos}) and the definition of the left moving part of the momentum, Eq. (\ref{momento}), we can obtain the correlation function after integrating over the initial conditions
\bea
&&\langle \Pi_L(x_1,t)\Pi_L(x_2,t) \rangle=\int_0^{\infty} \frac{dk}{\sqrt{2k}}k^2\nonumber\\
&&\times e^{-ik(x_1-x_2)e^{-\kappa\int_0^t\sigma(s)ds}-\kappa\int_0^t\sigma(s)ds} \coth{\frac{\beta k}{2}}, \label{corrmal}
\ea
with the usual definition $\beta=(k_B T_0)^{-1}$. 

Either for the region $-a<x<a$ where the velocity changes with position, or the regions where the velocity is homogeneous (as stated, upon the replacements $\kappa\mapsto 0$ and $\sigma \mapsto v \sigma$), the expression below depends only on $(x_1-x_2)$. Therefore, this expression cannot present the characteristic peak discussed. Accordingly, the regions where this solution is valid do not present the signature of the quantum Hawking effect, as explained in previous sections.
\begin{figure}[h]
\includegraphics[width=0.4\textwidth]{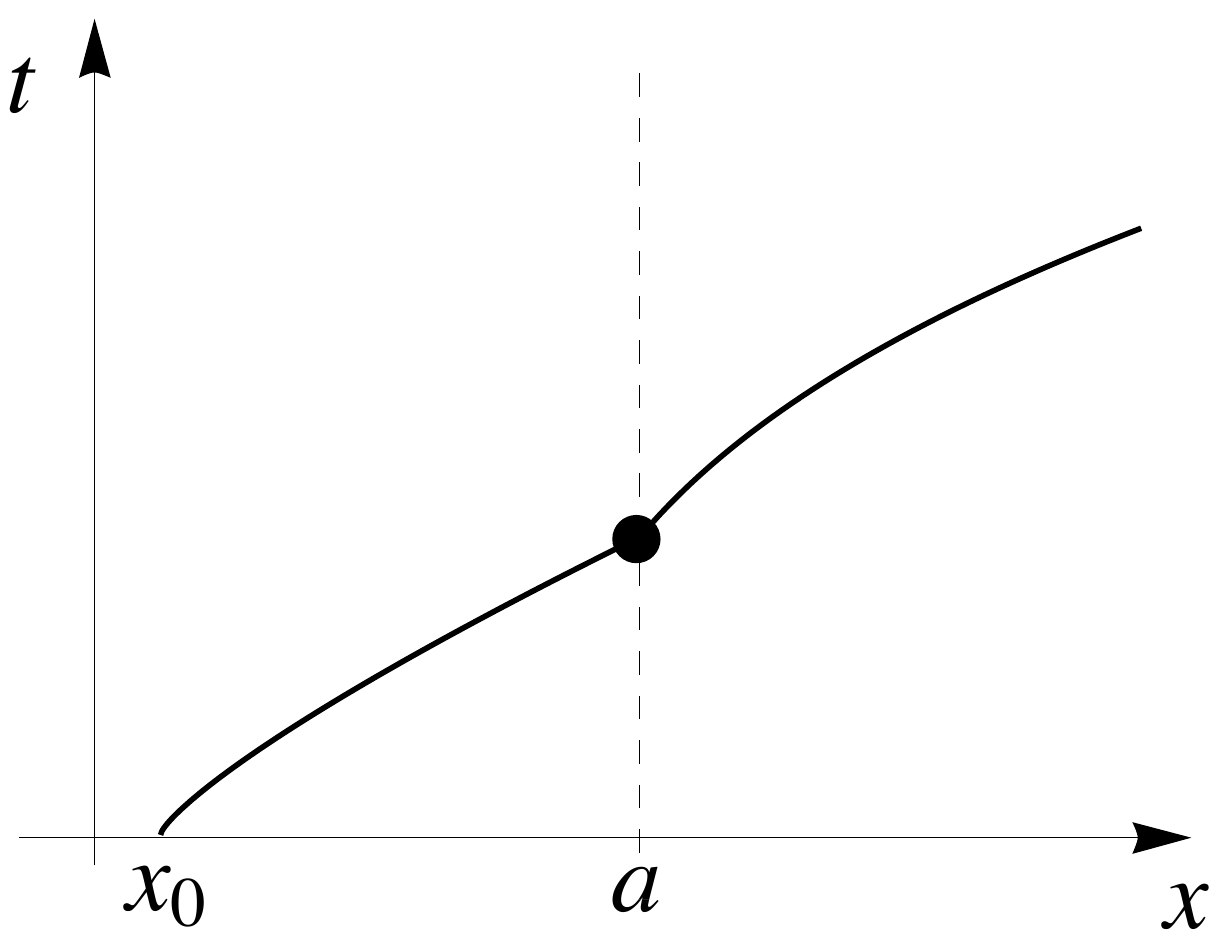}
\caption{\label{q}Schematic illustration of a characteristic curve that intersects the interface in $x=a$ and starts at $x_0$.}
\end{figure}

The correct way of getting the solutions of different regions is matching the characteristic curves as shown in Fig. \ref{q}. For example, if $x>a$ and different regions share the characteristic then $x_0$ is obtained matching the characteristic for $v(x,t)$ constant with the characteristic for $x<a$ and finally finding the $x_0$ corresponding to the latter. The signature of the entanglement between the Hawking pair was shown to be present in the left component of the correlation so we are only interested in the  curves associated with this modes. 

To carry on this procedure, we approximate for long times ($t\gg\tau$) the characteristic for $|x|<a$ as $x(t)\simeq x_0e^{\kappa t}$. In the region $x>a$, the characteristic is
\beq 
x(t)=A-t+f(t)v_{\rm max},
\eeq
where $A$ is a constant of integration, $f(t)\equiv\int_0^tds \sigma(s)$ and $v_{max}\approx1+\kappa a$. The relationship between $A$ and $x_0$ can be found recalling that
\beq
x(t_a)=a \;\;\; \Rightarrow ~~~t_a = \kappa^{-1} \log{a/x_0},
\eeq
because the $|x|<a$ characteristic reach $x=a$ at $t_a$ and
\beq
A-t_a+f(t_a)v_{\rm max}=a,
\eeq
because the $x>a$ characteristic must match the previous one at $x(t_a)=a$.

From this two equations, for times $t\gg\tau$, we can obtain $A$ as a function of $x_0$ and then use $x_0$ as a function of $x,t$ and put it in the $e^{ikx_0}$ factor, finding
\bea
x_0&=&ae^{(x+t-f(t)v_{\rm max}-a)/a}, ~~~~~\text{for }x>a\text{ and}\\
x_0&=&-a e^{-(x+t-f(t)v_{\rm min}-a)/a}, ~~~~\text{for }x<-a.
\ea

 Therefore the left moving modes are given by 
\bea
u_{Lk}(x)&=&\frac{1}{\sqrt{2|k|}} e^{ik a e^{(x+t-f(t)v_{\rm max}-a)/a}}~~~,\text{for }x>a \nonumber\\
u_{Lk}(x)&=&\frac{1}{\sqrt{2|k|}} e^{-ik a e^{(-x-t+f(t)v_{\rm min}+a)/a}}, \text{for }x<-a. \nonumber
\ea
\begin{figure}[h]
\includegraphics[width=0.98\columnwidth]{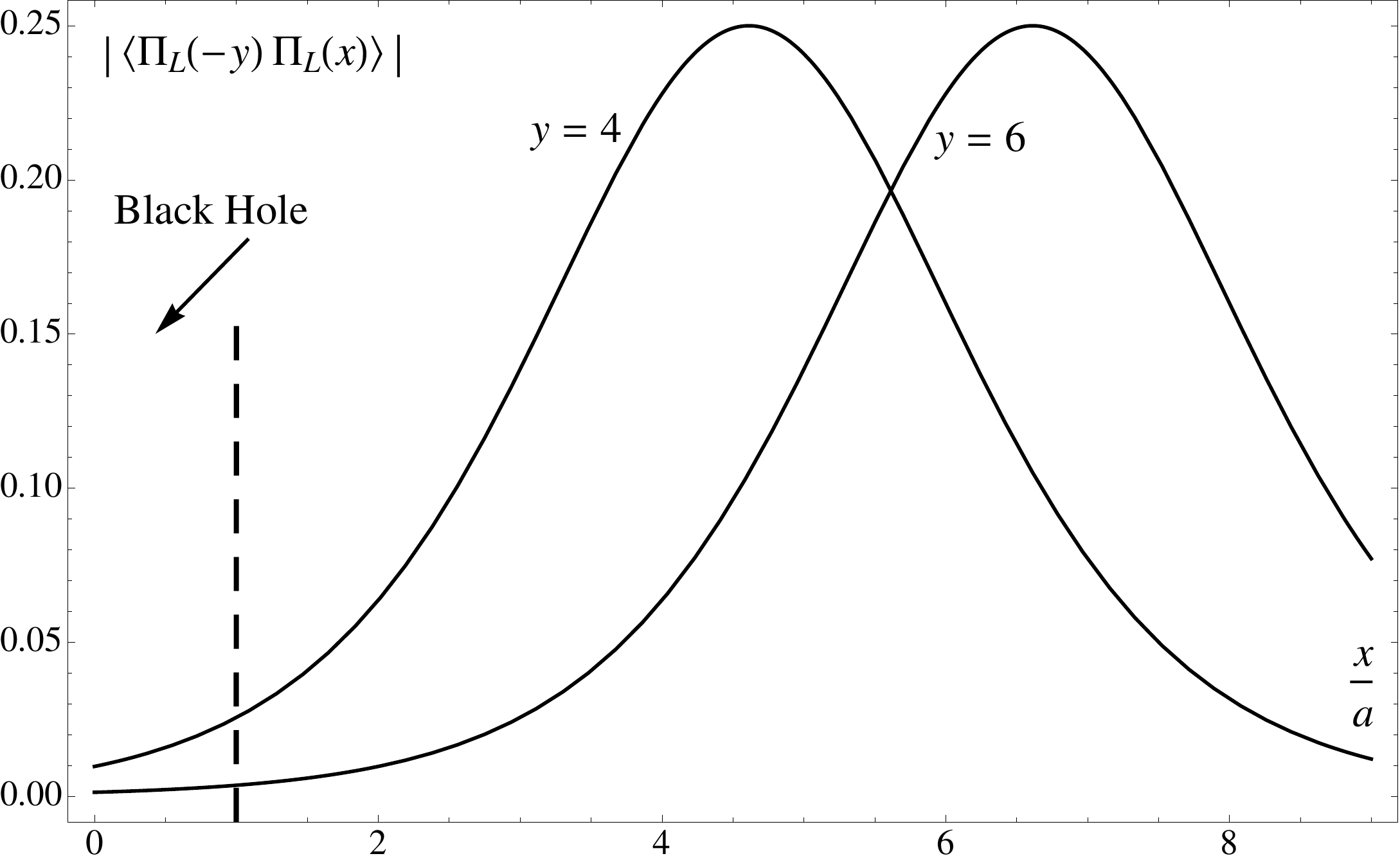} 
\caption{\label{grafico1} Plot of the absolute value of the left-moving component of the correlation function for long times $t=100\tau$. The maximum velocity is $1.1$, the minimum $0.9$ and the initial temperature is zero $T_0=0$.}
\end{figure}
Finally, we calculate $\langle \phi_L(x_1,t) \phi_L(x_2,t) \rangle$ for an initial thermal state at temperature $\beta=(k_BT_0)^{-1}$, and deriving with respect to $x,t$ according to Eq. (\ref{momento}) in order to obtain the momentum two point function. The final result is
\bea
\langle \Pi_L(x_1,t)&& \Pi_L(x_2,t)\rangle=-\frac{\pi^2}{\beta^2} e^{\frac{x_1-x_2}{a}}\big[\cosh{\big(\frac{t}{\tau}\big)}\big]^{-\frac{\Delta v\tau}{a}}\nonumber\\
&&\bigg\{{\rm cosech}\big[\frac{a\pi}{\beta}e^{-\frac{a+t+x_2}{a}}\big[\cosh{\big(\frac{t}{\tau}\big)}\big]^{\frac{v_{\rm max}\tau}{a}}\nonumber\\
&&\times\big(e^{\frac{2t+x_1+x_2}{a}}+e^{2}\big[\cosh{\big(\frac{t}{\tau}\big)}\big]^{\frac{\Sigma v\tau}{a}}\big)\big]\bigg\}^2, \label{corrrr}
\ea
where $\Delta/\Sigma v \equiv v_{\rm max}-/+v_{\rm min}$. This dependence of the correlation function between $y=-x_1=4,6$ and $x_2=x$ is shown in Fig. \ref{grafico1}. If $y$ is far from the event horizon, placed at $x=0$, the peak in the correlation will be far of the event horizon too, in the inverse direction. Even though the peak does not seems to modify with increasing $y$, it disappear for $y$ big enough, as we will see below. This can be concluded directly from the characteristic curves. If one follows the characteristic with $x_0=a$ then everything to the right will not be aware that there is an event horizon in the space; therefore, there should not be entanglement (peak in correlation) and there is none indeed. This is because in this region we have to use the solutions given by Eq. (\ref{corrmal}), which was shown not to present the signature.

Moreover, we can compare Fig. \ref{grafico1} of this paper with Fig. 1 of reference \cite{entre:unruh-cor}. We can conclude that for long times $t\gg\tau$, the one calculated for the black hole with the Israel-Hartle-Hawking coincide with the calculation in the present paper and in Ref. \cite{tesis}, that includes the collapse. Their behavior is the same, but little differences may arise from the fact that the velocity profile as a function of $x$ is different, since they use a smooth $\tanh(x/a)$ profile.
\begin{figure}[h]
\subfigure[]{
\includegraphics[width=0.9\columnwidth]{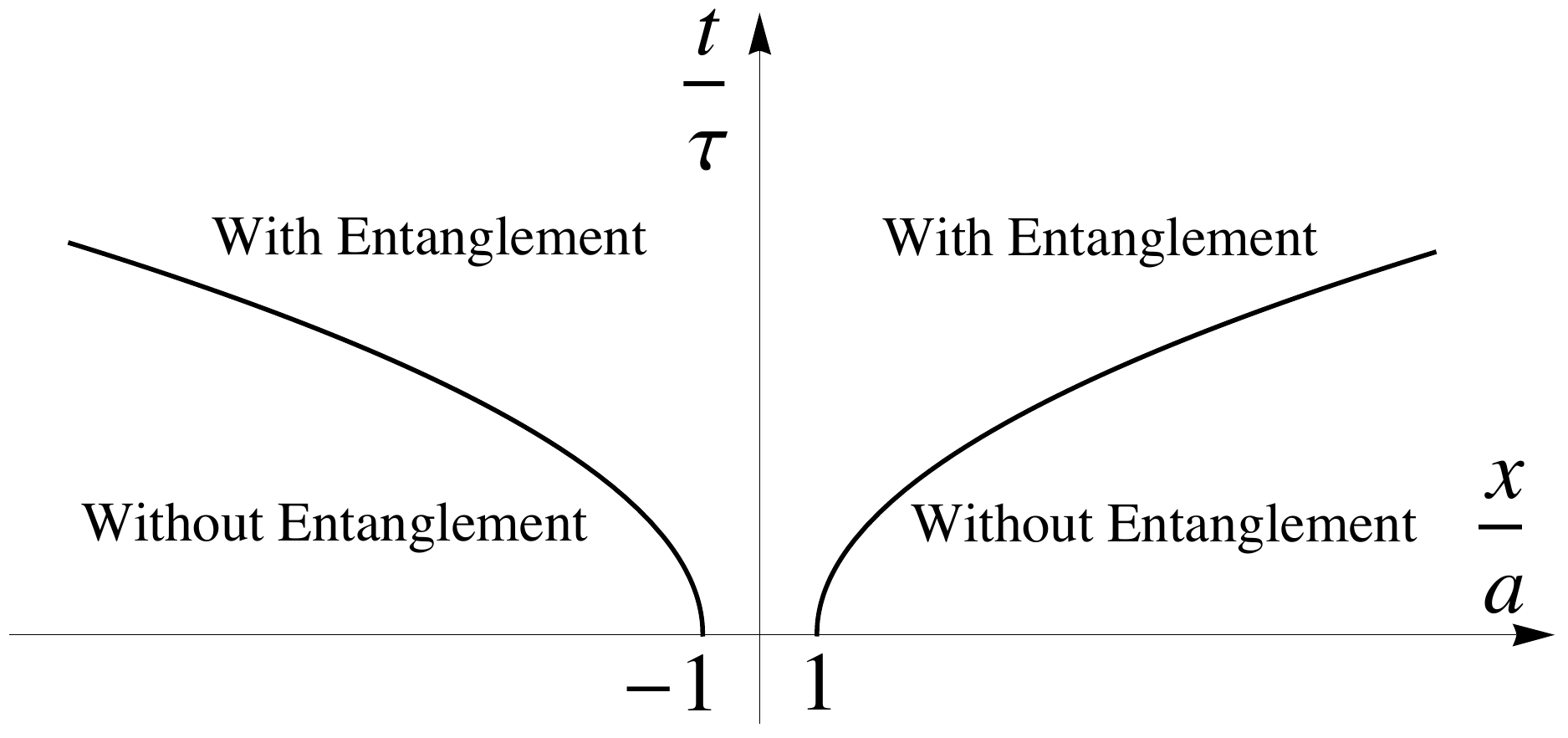}}
\subfigure[]{
\includegraphics[width=0.9\columnwidth]{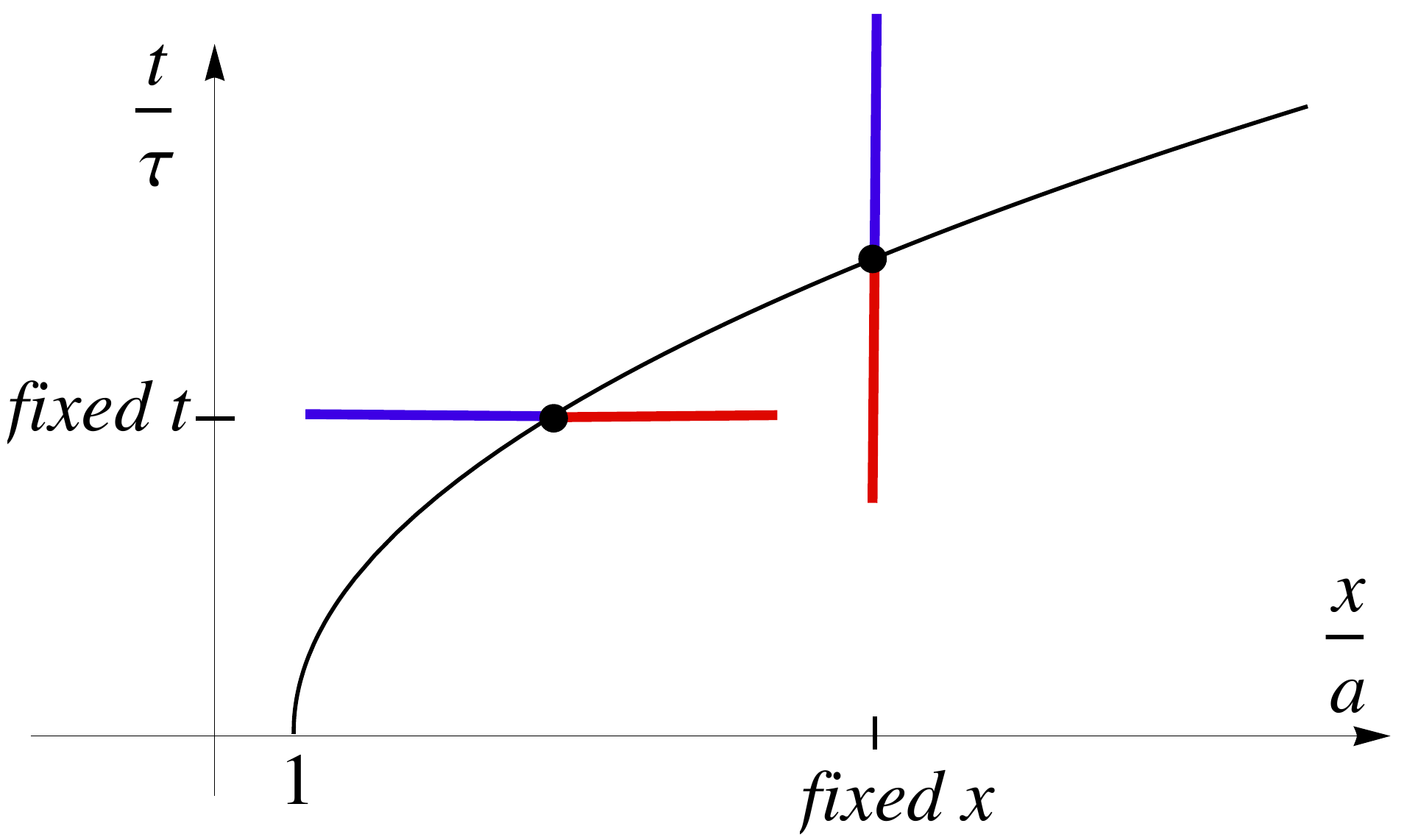}}
\caption{\label{borde}[Color Online] (a) Illustration of the regions where the peak of the correlation function is present, and therefore the entanglement between the Hawking pair particles and those that not. (b) Behavior for fixed time or position. Red: no correlation peak. Blue: presence of correlation peak.}
\end{figure}

Nonetheless, this formalism allows us also to study the evolution of the correlation function. As we explained, there are two special characteristic curves, that start at $x_0=\pm a$, and they separate two regions, one that presents the peak that reflects the entanglement and another one that does not. Therefore, the interface that separates both regions, which is shown in Fig. \ref{borde}, is given by 
\bea
x_{\pm}(t)&=&\pm a\bigg(1+\kappa  \int_0^t \sigma(t) ds\bigg)\nonumber\\
&=&\pm a \bigg(1+\kappa\tau \ln{\cosh{\frac{t}{\tau}}}\bigg).
\ea
As can be seen in Fig. \ref{borde}(b) if we study the correlation at fixed position $x$ there exist a definite time $t_E$ such that for $t<t_E$ there is no entanglement and it is generated later for $t>t_E$. This time can be found solving the equation $x_+(t_E)=x$.

 In turn, if we study the correlation at fixed time $t$, then there exists a position $x_E$ such that if $a<x<x_E$ there is no peak in the correlation and for $x>x_E$ it is present. 

For example, for fixed time $t=100\tau$ with the parameters of Fig. \ref{grafico1} one has the presence of the correlation peak only for $|x|\lesssim 11$, therefore it is correct to use the expression in Eq. (\ref{corrrr}) and not that of Eq. (\ref{corrmal}) to make the plot in Fig. \ref{grafico1}. The fact that this transitions are not smooth is due to the abrupt changes at $x=\pm a$ in the velocity profile.

These two features of the solution found can be compare with the numerical simulations done for the ion ring, \cite{cirac}. The fact that the correlation peak has a finite spatial extension can be clearly seen in Fig. 9 of Ref. \cite{cirac}, where this magnitude is calculated numerically. The fact the the entanglement needs a definite time to develop is also observed in Fig. 14 of Ref. \cite{cirac} through the numerical computation of the logarithmic negativity.
\begin{figure}[h]
\includegraphics[width=0.98\columnwidth]{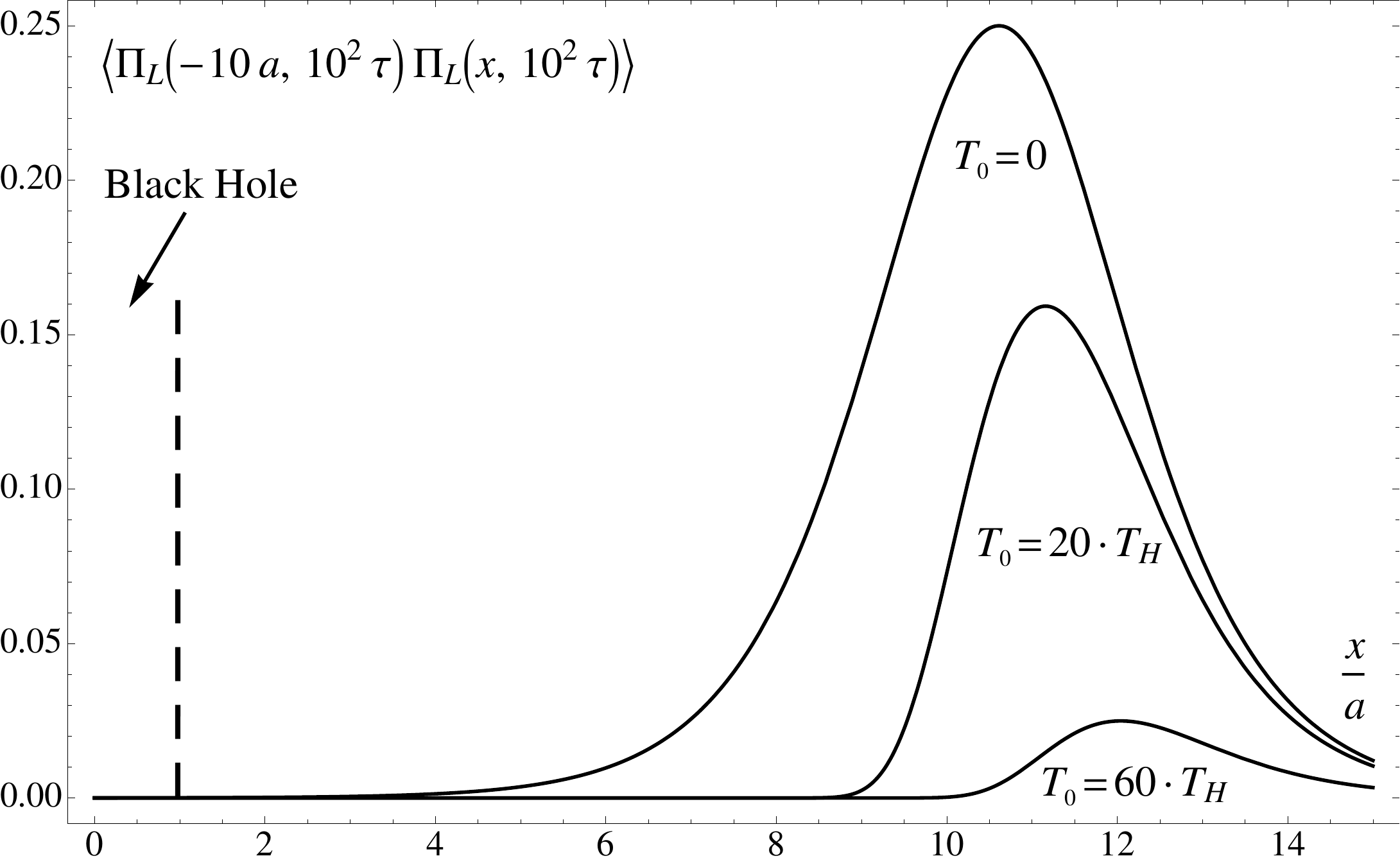} 
\caption{\label{grafico2} Plot of the absolute value of the correlation function of the momentum for long times $t=100\tau$. The maximum velocity is $1.1$, the minimum is $0.9$ and the initial temperatures are $T_0=0,20\cdot T_H, 60\cdot T_H$.}
\end{figure}

This formalism also allows us to study how the correlation, and therefore the entanglement, depends on the initial temperature of the field $T_0$, as can be seen in Eq. (\ref{corrrr}). In particular, our results should reproduce the behavior shown in Fig. 14 of Ref. \cite{cirac}, and the loss of entanglement of the Hawking pair at high temperatures. First we compute the Hawking temperature, given by (see Ref. \cite{unruhrob})
\beq
T_H=\frac{1}{2\pi}\frac{|v_{\rm max}-v_{\rm min}|}{2a}.
\eeq
In Fig. \ref{grafico2} we plot the dependence of the peak in the correlation as a function of the temperature in units of the Hawking temperature. We observe how the peak is diluted and therefore how the Hawking pair entanglement is lost for high enough initial temperatures.
\subsection{Entanglement: Open System.}
In this last section, we will try to estimate the correction introduced by the presence of an environment in the correlation function. Therefore, we will use the expression presented previously
\bea
\langle \phi_1\phi_2 \rangle&&= \langle \phi_{O1}\phi_{O2} \rangle_{\text{in}}- G_{2x'}{\bf D}_{x'x}\langle\phi_{Ox}\phi_{O2}\rangle_{\rm in}\nonumber\\
&&- G_{2x'}{\bf D}_{x'x}\langle\phi_{Ox}\phi_{O1}\rangle_{\rm in}+G_{1x'}{\bf N}_{x'x}G_{2x},
\ea
together with the solution for the modes found in the previous sections. Regarding the environment, we will use an ohmic bosonic bath with bilinear coupling. The dissipation kernel is given by
\beq
{\bf D}(s,s')=-\lambda^2 \frac{\partial}{\partial s} \delta(s-s'). \label{entre:dis}
\eeq

In this case, obtaining an expression analytic for $\lb \Pi_L\Pi_L\rb$ is more involved since the integrals must be performed over all possible regions and the final dependence of $k$ make the integrals more complicated. To solve the second inconvenience  we will study each mode separately, characterized by $k$. We will use the expression in Eq. (\ref{entre:dis}) to eliminate the integrals in $s'$, after integrating by parts the Dirac delta function. The remaining integrals can be approximated by the main contribution in the coincidence limit. Regarding the last term, for high temperatures the noise kernel is ${\bf N}(s_1,s_2)=\lambda^2T_0\delta(s_1-s_2)$ and the resulting integral is done using 
%\beq 
%\int_0^{t_1} ds\int  dx' G_{\text{ret}}(t_1,s|x_1,x')G_{\text{ret}}(t_2,s|x_2,x') =G_{\text{ret}}(t_2,t_1|x_2,x_1),
%\eeq 
basic properties of Green functions. After performing the derivatives necessary to obtain $\lb \Pi_L\Pi_L\rb$ we can estimate the correlation function at high temperatures. 
\begin{figure}[h]
\includegraphics[width=0.98\columnwidth]{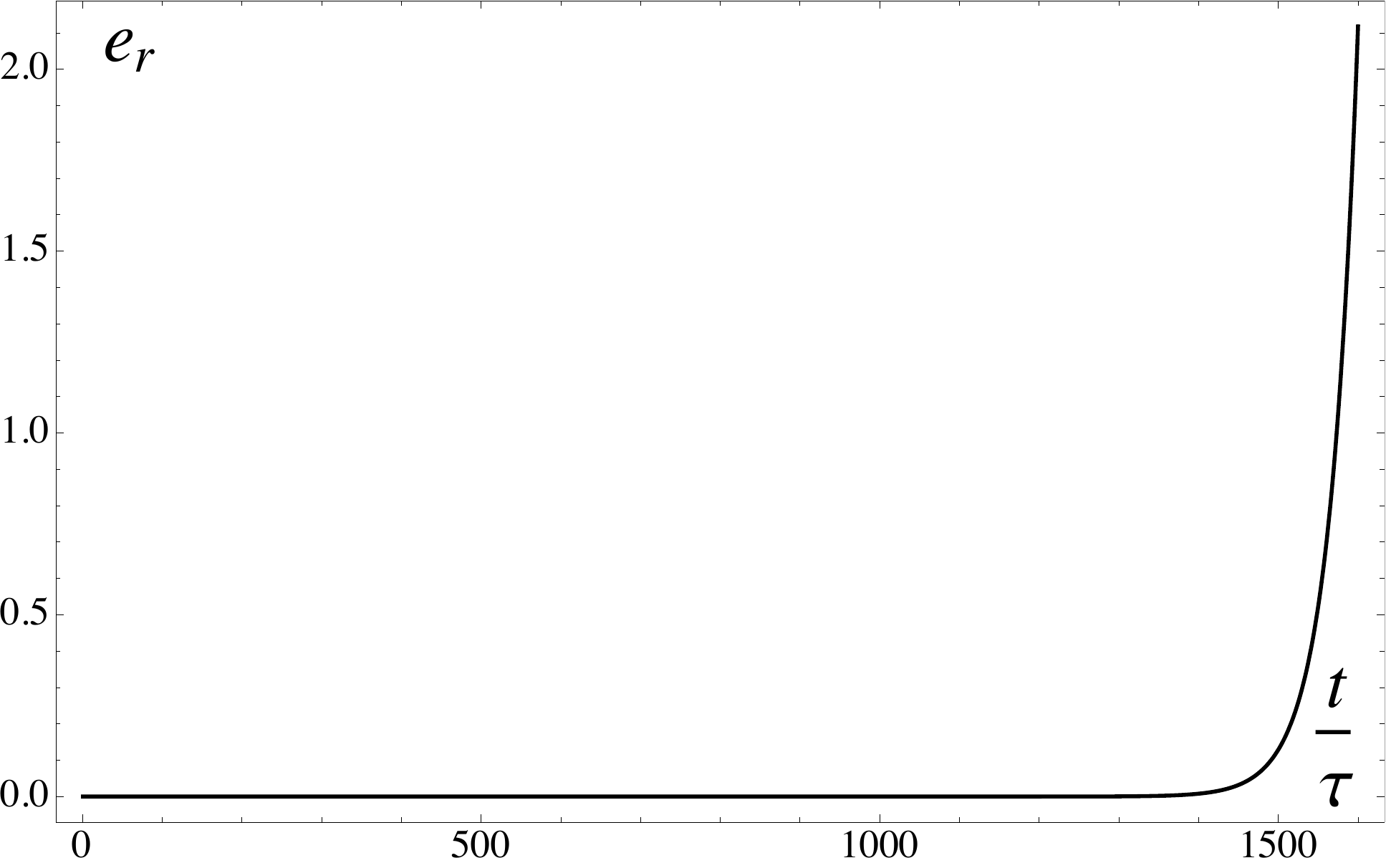} 
\caption{\label{grafico3} Relative contribution $e_r$ for high temperatures, $T\sim 100\cdot T_H$, evaluated on the position of the peak in the correlation function and for coupling $\lambda\sim 10^{-7}$. }
\end{figure}
Fig. \ref{grafico3} shows the result of the relative contribution of the term associated with the environment with respect to the solution for the closed black hole, given by
\beq
e_r\equiv \bigg|\frac{\lb \Pi_{L1}\Pi_{L2}\rb_C(k)-\lb \Pi_{L1}\Pi_{L2}\rb_O(k)}{\lb \Pi_{L1}\Pi_{L2}\rb_C(k)}\bigg|,
\eeq
where the $k$ dependence means that
\beq
\lb \Pi_{L1}\Pi_{L2}\rb=\int dk \lb \Pi_{L1}\Pi_{L2}\rb(k).
\eeq
This magnitude was computed for small $k$ since it decreases with increasing $k$. For small times the influence of the environment is negligible and the system behaves as an effectively closed system. For later times the relative contribution begins to grow. Even though we could not integrate to obtain the result as a function of $x,t$ we expect that this abrupt increase of the environmental influence produces decoherence and we take it as a signature of the loss of correlation and therefore, the entanglement between the Hawking pair particles.
\section{Closing Remarks} \label{sec:conc}
In this paper we studied the influence of an environment in acoustic black holes as an open quantum system. We used the circular ion trap setup but the analysis can be applied to any acoustic black hole, only details relevant to the environment could change. We put our attention to the process of decoherence induced by this environment in contradistinction with the quantum to classical transition considered previously induced by thermal equilibrium effects \cite{cirac}. As we explained in the Introduction, since the Hawking effect is purely quantum mechanical, then a quantum to classical transition would jeopardize the possible measurement of the radiation, therefore decoherence must be controlled in the experiment. 

We used the open quantum system approach to calculate de decoherence time of an acoustic black hole in presence of an environment for both zero and finite (but small) temperatures \cite{tesis}. Taking into account that the decoherence time is smaller than the approximate measurement time and worst, the ``collapse" time, improved parameters are needed and provided here to make the measurement possible. This analysis could be easily extended to other realizations of acoustic black hole such as BEC, moving dielectrics, waveguides, etc.

In order to achieve generality for this work to be interesting for any acoustic black hole, we made certain approximations. It would be interesting for future works to perform this derivation numerically taking as a starting point the exact dynamics of the particular acoustic black holes' setup. Moreover, regarding the environment one could consider more involved bosonic reservoirs that for example gets dragged to some extent by the moving ions instead of being fixed with respect to the laboratory.

We also provide a derivation of the correlation between the Hawking phonons as a function of time and temperature and we check that the relative contribution of the environment to this magnitude is irrelevant. The derivation included the transient due to the ``collapse" of the acoustic black hole and could be applied to the usual Schwarchild black hole as in \cite{entre:birrel-davies}.

 The main complication in this approach is the presence of boundaries between different velocity profiles, since the overall is not continuously differentiable. For future works, it would be interesting to find a smooth spatial velocity profile that presents an event horizon and allows to carry an analytic treatment to the end\footnote{For example, we tried to use a $\tanh{x/a}$ profile but exact analytical expressions for the characteristic curves could not be found.}, in order to obtain the correlation in the presence of an environment as a function of $t$ and $x$ instead of Fig. \ref{grafico3}. 
 
 Finally, another interesting extension of this work would be to study the quantum to classical transition due to the non linearities in the original hamiltonian, Eq. (\ref{calz}). The non equilibrium self interaction of the degrees of freedom  in this case (as opposed to the thermal equilibrium situation considered in Ref. \cite{cirac}) also introduces dissipation and noise in the same fashion as an environment and also induces decoherence. The decoherence time associated with this effect depends on intrinsic parameters of the system such as the mass, electric charge of the ions, etc.; as opposed to the external parameter $\gamma$ considered here. Therefore, this study would impose critical bounds on the realizability of the measurement of the Hawking radiation with any setup.
\section*{Acknowledgements}
This work was supported by UBA, CONICET and ANPCyT. GJT is also supported by CIN. The authors thank Esteban Calzetta, Juan Ignacio Cirac, Diana L\'opez Nacir and Juan Pablo Paz for useful discussions.
\appendix
\section{Diffusion coefficient at finite temperature} \label{A}
The dissipation kernel for finite temperature is defined as
\beq
D(t)=\frac{\tilde{\gamma}^2}{2} \int_0^{\infty} d\nu \int_0^t ds \nu \coth{\frac{\beta\hbar\nu}{2}} \cos{\nu s} \cos{\omega s}.
\eeq
As stated in section \ref{temtdec}, we will use the following approximation
\beq
\coth{\frac{\beta\hbar\nu}{2}} \approx \begin{cases}
\frac{2}{\beta\hbar\nu}    &  \text{si }\frac{\beta\hbar\nu}{2}\ll1\\
1   & \text{si }\frac{\beta\hbar\nu}{2}\gg1.
\end{cases}
\eeq
Having this in mind, the coefficient can be approximated integrating separately over the regions where the different approximations are valid, i.e.
\begin{eqnarray*}
2\tilde{\gamma}^{-2}D(t,\beta)&=&\bigg(\int_0^{2(\beta\hbar)^{-1}} d\nu+ \int_{2(\beta\hbar)^{-1}}^{\infty}d\nu \bigg) \nu \coth{\frac{\beta\hbar\nu}{2}}\\
&& \times \int_0^tds \cos{\omega s} \cos{\nu s}\\
&\approx&\int_0^{2(\beta\hbar)^{-1}} d\nu  \frac{2}{\beta\hbar} \int_0^tds \cos{\omega s} \cos{\nu s}\\ &&+\int_{2(\beta\hbar)^{-1}}^{\infty}d\nu  \nu \int_0^tds \cos{\omega s} \cos{\nu s}.
\end{eqnarray*}
The first integral is
\bea
&&\int_0^{2(\beta\hbar)^{-1}} d\nu  \frac{2}{\beta\hbar} \int_0^tds \cos{\omega s} \cos{\nu s} =\nn
&&= \frac{-\text{Si}[t(-\frac{2}{\beta\hbar}+\omega)]+\text{Si}[t(\frac{2}{\beta\hbar}+\omega)]}{\beta\hbar}.
\ea
For small $2(\beta\hbar)^{-1}$ we can approximate
\bea
-\text{Si}[t(-\frac{2}{\beta\hbar}+\omega)]+\text{Si}[t(\frac{2}{\beta\hbar}+\omega)]&\approx& \frac{d\text{Si}(x)}{dx}\bigg|_{x=\omega t} \frac{4t}{\beta\hbar}\nn
&=&\frac{4 \sin{\omega t}}{\omega\beta\hbar},\nonumber
\ea
and therefore
\begin{equation}
\int_0^{2(\beta\hbar)^{-1}} d\nu  \frac{2}{\beta\hbar} \int_0^tds \cos{\omega s} \cos{\nu s} \approx \frac{4 \sin{\omega t}}{\omega\beta^2\hbar^2}.
\end{equation}
Regarding the second term, we will use the following
\bea
 &&\int_{2(\beta\hbar)^{-1}}^{\infty}d\nu  \nu \int_0^tds \cos{\omega s} \cos{\nu s}=\nn
 &=& \int_{0}^{\infty}d\nu  \nu \int_0^tds \cos{\omega s} \cos{\nu s}\nn
 &&- \int_{0}^{2(\beta\hbar)^{-1}}d\nu  \nu \int_0^tds \cos{\omega s} \cos{\nu s}.
\ea
The first term was calculated for the $T_0=0$ case and with the result $\omega \text{Si}(\omega t)$. The second term is
\begin{equation}
\int_{0}^{2(\beta\hbar)^{-1}}d\nu  \nu \int_0^tds \cos{\omega s} \cos{\nu s} \approx  \frac{2 \sin{\omega t}}{\omega\beta^2\hbar^2}.
\end{equation}
Finally the new diffusion coefficient is given by
\begin{equation}
2\tilde{\gamma}^{-2}D(t,\beta)=\omega \text{Si}(\omega t)+\frac{2 \sin{\omega t}}{\omega\beta^2\hbar^2}+\mathcal{O}(\beta^{-3}). \label{aaa}
\end{equation}
When one takes $\omega=0$ and $t=0$ the correction to order $\sim \mathcal{O}(T_0^2)$ vanishes. Therefore, a thermal state of finite temperature does not affect the relationship between the parameters $\tilde{\gamma}$ y $\gamma$. 

The next step is just to integrate the Eq. (\ref{aaa}) in order to obtain the relevant diffusion coefficient.
\section{Modes of the Wave Equation}\label{solcar}
The equation we have to solve is Eq. (\ref{eq:aproxcoup}) which as noted above, can be written as $\partial_R\partial_L\phi=0$. This way, the more general solution is $\phi_O=\phi_L+\phi_R$, in terms of the left and right moving modes.

The first equation to solve is $\partial_L\phi_L=0$
\beq
\frac{\partial \phi_L}{\partial t}+(v(x,t)-1)\frac{\partial\phi_L}{\partial x}=0.
\eeq 
This equation is of the same form that the Eq. (\ref{ec:car}) with $b(x,t)=1$, $a(x,t)=v(x,t)-1=\sigma(t)\kappa x + \sigma(t)-1$ and $c(x,t)=0$. 

The definition of the characteristic curves is
\beq
\frac{dx}{ds}=\sigma(t)\kappa x+\sigma(t)-1 \;\;\;\;\;\;\ {\rm and}\;\;\;\;\;\;\;\;\; \frac{dt}{ds}=1.
\eeq
From the second equation one obtains $t=s$, where we impose that $t(0)=0$. Using this relationship in the first equation we get
\beq
\frac{dx}{dt}-\sigma(t)\kappa x=\sigma(t)-1.
\eeq
The solution of this equation is
\bea
x(t)&=&e^{\kappa\int_0^t\sigma(s)ds}\nn
&&\bigg(x_0- \int_0^t (1-\sigma(t))e^{-\kappa\int_0^s\sigma(s')ds'} ds\bigg),
\ea
where we used the condition $x(0)=x_0$. Then, $t(x_0,s)=t=s$ and inverting to get $x_0(x,t)$ we obtain
\bea
x_0(x,t)&=&xe^{-\kappa\int_0^t\sigma(s)ds}\nn
&&+ \int_0^t (1-\sigma(t))e^{-\kappa\int_0^s\sigma(s')ds'} ds.
\ea
The differential equation to solve for the evolution of the field along the curve is simply
\beq
\frac{d\phi}{ds}=0 \;\;\;\; \Rightarrow \; \phi(x(x_0,s),t(x_0,s))=\phi(x_0,0).
\eeq
Finally the solution is
\bea
\phi_L(x,t)&=&\phi\bigg(xe^{-\kappa\int_0^t\sigma(s)ds}\nn
&&+ \int_0^t (1-\sigma(t))e^{-\kappa\int_0^s\sigma(s')ds'} ds,0\bigg).
\ea
To rewrite it in a useful way, taking into account that we will have to eventually integrate over the initial conditions, we expand in modes $\phi(x,0)=\int dk/2\pi e^{ikx} \phi_k$, then the solution for the field is
\bea
\phi_L(x,t)&=&\int \frac{dk}{2\pi} \phi_k \text{exp}\bigg\{ik\bigg(xe^{-\kappa\int_0^t\sigma(s)ds}\nn
&&+ \int_0^t (1-\sigma(t))e^{-\kappa\int_0^s\sigma(s')ds'} ds\bigg)\bigg\}.
\ea

Having an expression for the left moving modes, we now have to solve the equation for the \emph{right} modes, $\partial_R\tilde{\phi}_R=0$,
\bea
\bigg(\frac{\partial}{\partial t}+ \frac{\partial}{\partial x}v(x,t) +\frac{\partial}{\partial x}\bigg)\tilde{\phi}_R&=&0\nonumber\\
\bigg(\frac{\partial}{\partial t} + \sigma(t)\kappa+\sigma(t)(1+\kappa x)\frac{\partial}{\partial x}+ \frac{\partial}{\partial x}\bigg)\tilde{\phi}_R&=&0\nonumber\\
\bigg(\frac{\partial}{\partial t} + \sigma(t)\kappa+(\sigma(t)+\sigma(t)\kappa x+1)\frac{\partial}{\partial x}\bigg)\tilde{\phi}_R&=&0.
\ea

Again,  $t=s$, but the spatial part of the characteristic is
\bea
x(t)&=&e^{\kappa\int_0^t\sigma(s)ds}\nn
&&\bigg(x_0+\int_0^t (\sigma(s)+1)e^{-\kappa\int_0^s\sigma(s')ds'} ds\bigg),
\ea
and after taking the inverse $x_0$
\bea
x_0(x,t)&=&xe^{-\kappa\int_0^t\sigma(s)ds}\nn
&&-\int_0^t (1+\sigma(s))e^{-\kappa\int_0^s\sigma(s')ds'} ds.
\ea
Now the equation for the value $\phi_R$ along the characteristics is
\beq
\frac{d\phi}{dt}=-\sigma(t)\kappa\phi,
\eeq
then
\beq
 \phi(x(x_0,s),t(x_0,s))=\phi(x_0,0) e^{-\kappa\int_0^s \sigma(s')ds'}.
\eeq
Therefore, expanding again in modes 
\bea
\tilde{\phi}_R&&(x,t)=\int \frac{dk}{2\pi} \tilde{\phi}_k \text{exp}\bigg\{ik\bigg(xe^{-\kappa\int_0^t\sigma(s)ds}\nn
&&-\int_0^t (1+\sigma(t))e^{-\kappa\int_0^s\sigma(s')ds'} ds\bigg)\nn
&&-\kappa\int_0^t \sigma(s)ds\bigg\}.
\ea
Taking advantage of this result for $\tilde{\phi}_R$, the right moving mode can be found from $(\partial_t+v(x,t)\partial_x-\partial_x)\phi_R=\tilde{\phi}_R$. To solve this equation we propose a solution of the form
\bea
\phi_R(x,t)&&=\int \frac{dk}{2\pi} \phi_k(x,t) \text{exp}\bigg\{ik\bigg(xe^{-\kappa\int_0^t\sigma(s)ds}\nn
&&+ \int_0^t (1-\sigma(s))e^{-\kappa\int_0^s\sigma(s')ds'} ds\bigg)\bigg\}.
\ea
If we use this result in the equation and use that the exponential is a solution of the homogeneous equation of the left moving modes, we obtain
\bea
\partial_L\phi_R&=&\int \frac{dk}{2\pi} \partial_L\phi_k(x,t){\rm exp}\bigg[ik\big(xe^{-\kappa\int_0^t\sigma(s)ds}\nn
&&+\int_0^t (1-\sigma(s))e^{-\kappa\int_0^s\sigma(s')ds'} ds\big)\bigg].
\ea
On the other hand, the solution $\tilde{\phi}_R$ can be rewritten as
\bea
\tilde{\phi}_R(x,t)&&=\int \frac{dk}{2\pi} \tilde{\phi}_k \text{exp}[-2ik\int_0^te^{-\kappa\int_0^s\sigma(s')ds'} ds-\kappa\int_0^t \sigma(s)ds]\nn
&&\times e^{ik\big(xe^{-\kappa\int_0^t\sigma(s)ds}+\int_0^t (1-\sigma(s))e^{-\kappa\int_0^s\sigma(s')ds'} ds\big)}.
\ea
Equating both sides of this equation gives
\beq 
\partial_L\phi_k(x,t)=\tilde{\phi}_k e^{-2ik \int_0^te^{-\kappa\int_0^s\sigma(s')ds'} ds-\kappa\int_0^t \sigma(s)ds}.
\eeq
To find the particular solution we can use a $\phi_k$ such that $\phi_k(x,t)=\Psi_k(t)$ and therefore
\beq
\partial_t\phi_k(t)=\tilde{\phi}_k e^{-2ik \int_0^te^{-\kappa\int_0^s\sigma(s')ds'} ds-\kappa\int_0^t \sigma(s)ds}
\eeq
Integrating in time, the solution is
\bea
\phi_R(x,t)&&=\int \frac{dk}{2\pi} \Phi_k e^{ik\big(xe^{-\kappa\int_0^t\sigma(s)ds}+ \int_0^t (1-\sigma(s))e^{-\kappa\int_0^s\sigma(s')ds'} ds\big)}\nn
&& \int_0^tds e^{-2ik \int_0^{s}e^{-\kappa\int_0^{s'}\sigma(s'')ds''} ds'-\kappa\int_0^s \sigma(s')ds'}.
\ea
The remaining steps are just to write down the full solution and cast it into the usual mode expansion form. To do this one has to write $\phi_{kL}$ and $\phi_{kR}$ in terms of the Fourier transform of the initial condition of the field and the conjugate momentum $\phi_k(t=0)$ and $\Pi_k(t=0)$.


\begin{thebibliography}{5}
\bibitem{hawking}
S. Hawking,
Nature (London) {\bf 248}, 30 (1974); 
Commun. Math. Phys. {\bf 43}, 199 (1975);
J. Hartle \& S. Hawking,
Phys. Rev. D {\bf 13}, 2188 (1976).

\bibitem{unruh-prl}
W.G.~Unruh,
%{\em Experimental Black Hole Evaporation?},
Phys. Rev. Lett. {\bf 46}, 1351 (1981).

\bibitem{bec}
L. J. Garay, J. R. Anglin, J. I. Cirac, and P. Zoller,
%{\em Sonic Analog of Gravitational Black Holes in Bose-Einstein Condensates},
Phys. Rev. Lett. {\bf 85}, 4643 (2000);
%L.~J.~Garay, J.~R.~Anglin, J.~I.~Cirac, and P.~Zoller,
%{\em Sonic black holes in dilute Bose-Einstein condensates},
Phys. Rev. A {\bf 63}, 023611 (2001);
L.J.~Garay,
%{\em Black Holes In Bose-Einstein Condensates},
Int. J. Theor. Phys. {\bf 41}, 2073 (2002);
C. Barcelo, S. Liberati, and M. Visser,
%{\em Analogue gravity from Bose-Einstein condensates},
Classical Quantum Gravity {\bf 18}, 1137 (2001).

\bibitem{dielectric}
R. Sch\"utzhold, G. Plunien, and G. Soff,
%{\em Dielectric black hole analogs},
Phys. Rev. Lett. {\bf 88}, 061101 (2002);
I.~Brevik and G.~Halnes,
%{\em Effective Potential for Light in Moving Media},
%{\em Light rays at optical black holes in moving media},
Phys. Rev. D {\bf 65}, 024005 (2001).

\bibitem{nonlinear}
R. Sch\"utzhold and W. G. Unruh,
%{\em waveguides},
Phys. Rev. Lett. {\bf 95}, 031301 (2005).

\bibitem{not-slow}
W. G. Unruh and R. Sch\"utzhold,
%{\em On Slow Light as a Black Hole Analogue},
Phys. Rev. D {\bf 68}, 024008 (2003).

\bibitem{quilombo}
F. Belgiorno, S.L. Cacciatori, M. Clerici, V. Gorini, G.
Ortenzi, L. Rizzi, E. Rubino, V. G. Sala, and D. Faccio,
Phys. Rev. Lett. {\bf 105}, 203901(2010);
R. Sch\"utzhold and W. G. Unruh,
\emph{ibid}. {\bf 107}, 149401 (2011);
F. Belgiorno, S. L. Cacciatori, M. Clerici, V. Gorini, G. Ortenzi, L. Rizzi, E. Rubino, V. G. Sala, and D. Faccio,
\emph{ibid}. {\bf 107}, 149402 (2011).

\bibitem{cirac}
B. Horstmann, B. Reznik, S. Fagnocchi, and J.I. Cirac,
%{\em waveguides},
Phys. Rev. Lett. {\bf 104}, 250403 (2010);
New J. Phys. {\bf 13}, 045008 (2011).

\bibitem{corrbec}
R.~Balbinot, A.~Fabbri, S.~Fagnocchi, A.~Recati and I.~Carusotto,
   %``Non-local density correlations as signal of Hawking radiation in BEC
   %acoustic black holes,''
   Phys.\ Rev.\  A {\bf 78}, 021603 (2008);
   %[arXiv:0711.4520 [cond-mat.other]].
   %%CITATION = PHRVA,A78,021603;%%
I.~Carusotto, S.~Fagnocchi, A.~Recati, R.~Balbinot and A.~Fabbri,
   %``Numerical observation of Hawking radiation from acoustic black holes in
   %atomic BECs,''
   New J.\ Phys.\  {\bf 10}, 103001 (2008);
   %[arXiv:0803.0507 [cond-mat.other]].
   %%CITATION = NJOPF,10,103001;%%
J.~Macher and R.~Parentani,
   %``Black-hole radiation in Bose-Einstein condensates,''
Phys.\ Rev.\ A {\bf 80}, 043601 (2009).
%   arXiv:0905.3634 [cond-mat.quant-gas].
   %%CITATION = ARXIV:0905.3634;%%

\bibitem{medicion}
C. Monroe, D. M. Meekhof, B. E. King, S. R. Jefferts, W.M. Itano, and D.J. Wineland,
Phys. Rev. Lett. {\bf 75}, 4011 (1995);
D.J. Wineland \emph{et al.}
J. Res. Natl. Inst. Stand. Technol. {\bf 103}, 259 (1998);
R. Sch\"utzhold, 
Phys. Rev. Lett. {\bf 97}, 190405 (2006).

\bibitem{tesis}
F. C. Lombardo and G. J. Turiaci, Phys. Rev. Lett. {\bf 108}, 261301 (2012).

\bibitem{unruhrob}
W. G. Unruh, Phys. Rev. D {\bf 51}, 2827 (1995).

\bibitem{entre:unruh-cor}
R. Sch\"utzhold and W.G. Unruh,
%{\em Quantum correlations across the black hole horizon},
Phys. Rev. D {\bf 81}, 124033 (2010).

\bibitem{tdec:wineland}
S. Schneider and G. J. Milburn,
Phys. Rev. A {\bf 59}, 3766 (1999);
C. J. Wyatt, B. E. King, Q. A. Turchette, C. A. Sackett, D. Kielpinski, W.M. Itano, C. Monroe, and D.J. Wineland,
Nature (London) {\bf 403}, 269 (2000);
Seidelin \emph{et al.},
%{\em waveguides},
Phys. Rev. Lett. {\bf 96}, 253003 (2006);
%Chiaverini \emph{et al.},
%Quant. Inf. \& Comp., {\bf 5}, 419 (2005).

\bibitem{tdec:entorno}
B. L. Hu, J. P. Paz, and Y. Zhang,
Phys. Rev. D {\bf 47}, 1576 (1993); 
J. P. Paz, S. Habib, and W. H. Zurek, 
\textit{ibid.} {\bf 47}, 488 (1993); 
F.C. Lombardo and P.I. Villar, 
Phys. Lett. A {\bf 371}, 190 (2007).
\bibitem{miles}
W.G. Unruh and R. Sch\"utzhold,
Phys. Rev. D {\bf 71}, 024028 (2005).

\bibitem{estadoinicial}
L.D. Romero and J.P. Paz, Phys. Rev. A {\bf 55}, 4070 (1997)
\bibitem{lombardo}
F.C. Lombardo and F.D. Mazzitelli,
Phys. Rev. D {\bf 53}, 2001(1996);
F. C. Lombardo, F. D. Mazzitelli, and R. J. Rivers,
Nucl. Phys. {\bf B672}, 462 (2003);
F.C. Lombardo and D.L\'opez Nacir,
Phys.Rev.D {\bf 72}, 063506 (2005). 

\bibitem{calhu}
E.A. Calzetta and B.L. Hu,
\emph{Nonequilibrium Quantum Field Theory},
(Cambridge University Press, Cambridge, England, 2008).

\bibitem{paula}
F.C. Lombardo and P.I. Villar,
Phys. Lett. A {\bf 336}, 16 (2005).

%\bibitem{entre:hartle-hawking}
%J. Hartle and S. Hawking,
%Phys. Rev. D {\bf 13}, 2188 (1976).

\bibitem{Boulware}
D.~G.~Boulware,
%{\em Quantum Field Theory in Schwarzschild and Rindler Spaces},
Phys.\ Rev.\ D {\bf 11}, 1404 (1975).
\bibitem{entre:birrel-davies}
N.D. Birrell and P.C.W. Davies,
\emph{Quantum Field Theory in Curved Space}
Cambridge Monographs, England, 1984.
\bibitem{IHH}
W.~Israel,
%{\em Thermo Field Dynamics of Black Holes},
Phys.\ Lett.\ A {\bf 57}, 107 (1976);


\bibitem{calzetta}
E.A. Calzetta, A. Roura, and E. Verdaguer,
%{\em Stochastic description for open quantum systems},
Physica (Amsterdam) {\bf 319A}, 188-212 (2003).

\bibitem{fuidos}
L.D Landau and E.M. Lifshitz, \emph{Fluid Mechanics}, Pergamon, 1959;
A.H. Shapiro,
{\em The Dynamics and Thermodynamics of Compressible Fluid Flow.}
(Ronald, New York (1953).

\end{thebibliography}
\end{document}